\documentclass[12pt]{article}
\usepackage[a4paper, top= 5cm,bottom=4cm, left=1.5 in, right=1.5 in]{geometry}
\usepackage{amsmath}
\usepackage{graphicx}
\usepackage{enumerate}
\usepackage{natbib}
\usepackage{url} 
\usepackage{setspace}

\usepackage{xr}
\makeatletter

\newcommand*{\addFileDependency}[1]{
	\typeout{(#1)}
	\@addtofilelist{#1}
	\IfFileExists{#1}{}{\typeout{No file #1.}}
}\makeatother

\usepackage{hyperref}
\hypersetup{
    colorlinks=true,
    linkcolor=black,
    filecolor=magenta,      
    urlcolor=cyan,
    citecolor = blue,
    pdftitle={Overleaf Example},
    pdfpagemode=FullScreen,
    }

\RequirePackage{mathrsfs, amsthm, amsmath, amsfonts, amssymb}
\usepackage{titlesec}
\renewcommand{\arraystretch}{1.9}
\usepackage{amsmath, amssymb, amscd, amsthm, amsfonts}
\usepackage{graphicx}
\usepackage{xr}
\usepackage{float}
\usepackage{color}
\usepackage{bm}
\usepackage{diagbox}
\usepackage{multirow}
\usepackage{makecell}
\usepackage{xcolor}
\usepackage{algorithm}
\usepackage{algpseudocode}
\usepackage{float}
\usepackage{tabularx}
\usepackage{ulem}
\usepackage{tabu}
\usepackage{graphicx}
\usepackage{subcaption}
\usepackage{hyperref}

\usepackage{setspace}
\begin{document}
	\date{}
	
\title{\bf Functional Clustering for Longitudinal Associations between Social Determinants of Health and Stroke Mortality in the US}
\author{
Fangzhi Luo\footnote{Department of Epidemiology and Biostatistics, College of Public Health, 	University of Georgia, Athens, GA, USA. The authors gratefully acknowledge \textit{the National Natural Science Foundation of China (grants nos. 12292980, 12292984 and 12231017) and the MOE project of key research institute of humanities and social sciences (grant no. 22JJD910001).}} \footnote{Jianbin Tan is the co-first author.}, ~
Jianbin Tan\footnote{Department of Biostatistics and Bioinformatics, School of Medicine, Duke University, Durham, NC, USA.}, ~
Donglan Zhang\footnote{Department of Foundations of Medicine, NYU Grossman Long Island School of Medicine, Mineola, NY, USA.},~
Hui Huang\footnote{Center for Applied Statistics and School of Statistics, Renmin University of China, Beijing, China}~ and~
Ye Shen\footnote{Department of Epidemiology and Biostatistics, College of Public Health, 
University of Georgia, Athens, GA, USA.}
}
    \maketitle
	
    \def\spacingset#1{\renewcommand{\baselinestretch}%
		{#1}\small\normalsize}
	
	
	
	\begin{abstract}
		Understanding the longitudinally changing associations between Social Determinants
		of Health (SDOH) and stroke mortality is essential for effective stroke
		management. Previous studies have uncovered significant regional disparities
		in the associations between SDOH and stroke mortality. However, existing
		studies have not utilized longitudinal associations to develop data-driven
		methods for regional division in stroke control. To fill this gap, we propose
		a novel clustering method to analyze SDOH-stroke mortality associations
		across U.S. counties. To enhance the interpretability of the clustering
		outcomes, we introduce a novel regularized expectation-maximization algorithm
		equipped with sparsity- and smoothness-pursued penalties, aiming at simultaneous
		clustering and variable selection in longitudinal associations. As a result,
		we can identify key SDOH that contributes to longitudinal changes in stroke
		mortality. This facilitates the clustering of U.S. counties based on the
		associations between these SDOH and stroke mortality. The effectiveness
		of our proposed method is demonstrated through extensive numerical studies.
		By applying our method to longitudinal data on SDOH and stroke mortality
		at the county level, we identify 18 important SDOH for stroke mortality
		and divide the U.S. counties into two clusters based on these selected
		SDOH. Our findings unveil complex regional heterogeneity in the longitudinal
		associations between SDOH and stroke mortality, providing valuable insights
		into region-specific SDOH adjustments for mitigating stroke mortality.
	\end{abstract}
	
	\noindent%
	\vfill
	
	\spacingset{1.5} 
	
	\section{Introduction}
	\label{sec: intro}

	The burden of stroke in the United States is enormous. As one of the most
	prevalent cardiovascular diseases, stroke remains consistently among the
	top five causes of death in the country (\citet{koton2014stroke}), leading
	to over 130{,}000 fatalities annually (\citet{holloway2014palliative}). In
	the effort to prevent stroke deaths, Social Determinants of Health (SDOH)---encompassing
	economic, social, and environmental conditions where people live, learn,
	work, and play (\citet{havranek2015social})---have garnered significant attention
	due to their strong associations with stroke mortality
	(\citet{reshetnyak2020impact,skolarus2020considerations,powell2022social}).
	In 2013, Centers for Disease Control and Prevention (CDC) released an important
	report on health disparities (\citet{frieden2013cdc}), where the relationships
	between SDOH and stroke mortality were extensively documented. In 2015,
	the impact of SDOH on stroke mortality was highlighted in an important
	scientific statement by the American Heart Association (AHA)
	(\citet{mozaffarian2015heart}). Since 2019, the AHA has consistently emphasized
	the importance of SDOH in its Annual Heart Disease and Stroke Statistics
	Report, spanning across all chapters (\citet{benjamin2019heart}). These declarations
	underscore the urgency of understanding the associations between SDOH and
	stroke mortality. This understanding is crucial for informing targeted
	interventions aimed at controlling stroke mortality by addressing underlying
	social, economic, and environmental factors.

	Recent findings have revealed a clear regional disparity in the associations
	between SDOH and stroke mortality. For instance,
	\citet{zelko2023geographically} identified statewise differences in these
	associations. Additionally, \citet{villablanca2016outcomes} and
	\citet{son2023social} observed significant disparities in the associations
	between rural and urban areas. To address such regional disparities, various
	statewise (\citet{gebreab2015geographic}) and rural-urban strategies
	(\citet{labarthe2014public,record2015community,kapral2019rural}) have been
	developed for region-specific stroke management. However, the existing
	literature primarily adopts prespecified regional division strategies when
	analyzing regional disparities, which do not guarantee that areas within
	a divided region share similar associations between SDOH and stroke mortality.
	This can potentially lead to ineffective regional division strategies for
	stroke control. Moreover, many SDOH exhibit longitudinal changes in their
	associations with stroke mortality
	(\citet{he2021trends,imoisili2024prevalence}), a factor crucial for timely
	policymaking in stroke management. Targeting these longitudinal associations,
	clustering becomes a vital tool for determining reasonable regional divisions
	in stroke prevention. Nevertheless, this remains an unsolved issue that
	requires further investigation.
	
	In this work, we propose a novel method for clustering longitudinal associations
	between SDOH and stroke mortality. Generally, the clustering is implemented
	based on the similarity among SDOH-stroke mortality associations across
	counties in the U.S., and the resulting clustering outcomes can inform
	region-specific prevention measures for controlling stroke mortality. To
	achieve this, we utilize county-level longitudinal data on SDOH and stroke
	mortality in the U.S., treating the longitudinally observed data in each
	county as functional data
	(\citet{RamsayJ.O.JamesO.2005Fda, wang2016functional}). Consequently, the
	concurrent associations between SDOH and stroke mortality are inherently
	functional objects, which can be effectively modeled using functional regression
	models
	(\citet{yao2005functional,wang2008variable,zhu2012multivariate, morris2015functional, kong2016partially, zhou2023functional}),
	where the coefficients are also functional, capturing the relationship
	between SDOH and stroke mortality over time. For the clustering process,
	we model these functional coefficients using a finite mixture model, leading
	to a finite mixture of functional regression models. This approach enables
	the clustering of U.S. counties into different regions based on the relationship
	between their SDOH and stroke mortality.
	
	\paragraph*{\textbf{Related Literature}} 
	Our task essentially differs from the common clustering procedures for
	longitudinal data
	(\citet{chiou2007functional,peng2008distance,mcnicholas2010model,jacques2013funclust,komarek2013clustering,jacques2014model,tang2016mixture}),
	which primarily apply to observable longitudinal outcomes. In contrast,
	the county-level longitudinal associations between SDOH and stroke mortality
	in our study are unobservable. To cluster these associations, one might
	consider applying finite mixtures of regression models
	(\citet{peel2000finite, khalili2007variable,khalili2013regularization}) to
	the SDOH and stroke mortality longitudinal data. However, this approach
	does not account for longitudinal changes in the associations within the
	clustering procedures, which could lead to unreliable clustering outcomes
	due to the disregard of longitudinal signals.
	
	In light of functional data analysis
	(\citet{ramsay2002applied, RamsayJ.O.JamesO.2005Fda,wang2016functional}),
	some studies proposed a finite mixture of functional regression models
	(\citet{yao2011functional, lu2012finite, wang2023functional}) to account
	for time-varying changes within mixture regression models.
	However, these methods do not address the issue of collinearity among covariates
	within their regression models. In our case, the collection of SDOH served
	as functional covariates, exhibits significant collinearity due to their
	derivation from multiple domains (e.g., social, economic, and environmental
	domains). Ignoring collinearity among the SDOH data may lead to misspecification
	in the functional regression model, potentially compromising the accuracy
	of the resultant clustering outcome.
	
	To enhance interpretability, existing studies analyzing SDOH-stroke mortality
	associations often adopt a preselection step on the SDOH covariates
	(\citet{tsao2022heart,tsao2023heart}), as their number is usually large,
	containing hundreds of variables. Without a proper selection of SDOH covariates,
	direct clustering using the existing methods
	(\citet{yao2011functional, lu2012finite, wang2023functional}) may be statistically
	inefficient due to the complex structure of longitudinal SDOH data, which confronts high-dimensional challenges and complex functional structures simultaneously. To
	solve this issue, one may perform variable selection in functional regression
	models
	(\citet{wang2008variable,goldsmith2017variable,ghosal2020variable,aneiros2022variable,cai2022variable})
	prior to clustering. However, these methods typically assume that associations
	between covariates and responses are invariant across samples, an assumption
	that limits their applicability to cases with heterogeneous associations.
	Moreover, performing selections of SDOH prior to clustering may also introduce
	biases for the subsequent clustering outcome, as the selection of SDOH
	is unrelated to the clustering process. To overcome theses challenges,
	it would be beneficial to connect the variable selection of functional
	covariates to the clustering of longitudinal associations; yet this topic
	is rarely discussed in the literature.
	
	\paragraph*{\textbf{Contributions}}
	In this article, we introduce a novel method for simultaneous clustering
	and variable selection of longitudinal associations between Social Determinants
	of Health (SDOH) and stroke mortality across U.S. counties. Our approach
	utilizes a Finite Mixture of Functional Linear Concurrent Models (FMFLCM),
	incorporating a new class of penalties that pursue both smoothness and
	sparsity. These penalties are designed to borrow information across different
	clusters within the functional regression models, enhancing statistical
	efficiency for both clustering and variable selection while addressing
	collinearity among functional covariates. For estimation, we develop a
	novel Regularized Expectation-Maximization (REM) algorithm that integrates
	the proposed penalties. This method facilitates the clustering of longitudinal
	associations while embedding a variable selection step for functional covariates.
	As a result, we can iteratively update the cluster memberships and the
	selected covariates within the REM algorithm, mitigating biases that may
	arise from selected covariates during the clustering process. Through this
	methodology, we provide a new data-driven approach for county-level regional
	division, aiming to deliver insights into region-specific stroke prevention
	measures for the U.S. counties.
	
	\paragraph*{\textbf{Organization}} 
	The remainder of this article is organized as follows: In Section~\ref{sec:method} we introduce the SDOH and stroke mortality dataset and
	then proceed to describe some of their data features in Section~\ref{sec: data source}. Subsequently, we present the Finite Mixture of
	Functional Linear Concurrent Models (FMFLCM) in Section~\ref{sec: flr}, followed by a demonstration of the sparsity- and smoothness-pursued
	penalties in Section~\ref{sec: penalty}, as well as the Regularized Expectation-Maximization
	(REM) algorithm in Section~\ref{sec: EM}. In Section~\ref{sec: simu} we
	conduct simulation studies to compare the proposed methods with competing
	approaches in terms of clustering performance, variable selection, and
	parameter estimation. In Section~\ref{sec: realdata} we apply the proposed
	clustering method to our dataset and present the clustering results, alongside
	the variable selection for the SDOH covariates. We conclude with a discussion
	in Section~\ref{sec: discussion}. The codes and datasets are publicly available
	at
	\url{https://github.com/fl81224/Functional-Clustering-of-Longitudinal-Associations}.\vadjust{\goodbreak}
	
	\section{Methodologies}\label{sec:method}
	\subsection{Data Source}
	\label{sec: data source}
	
	\begin{figure}[ht]
		\centering
		\includegraphics[scale = 0.4]{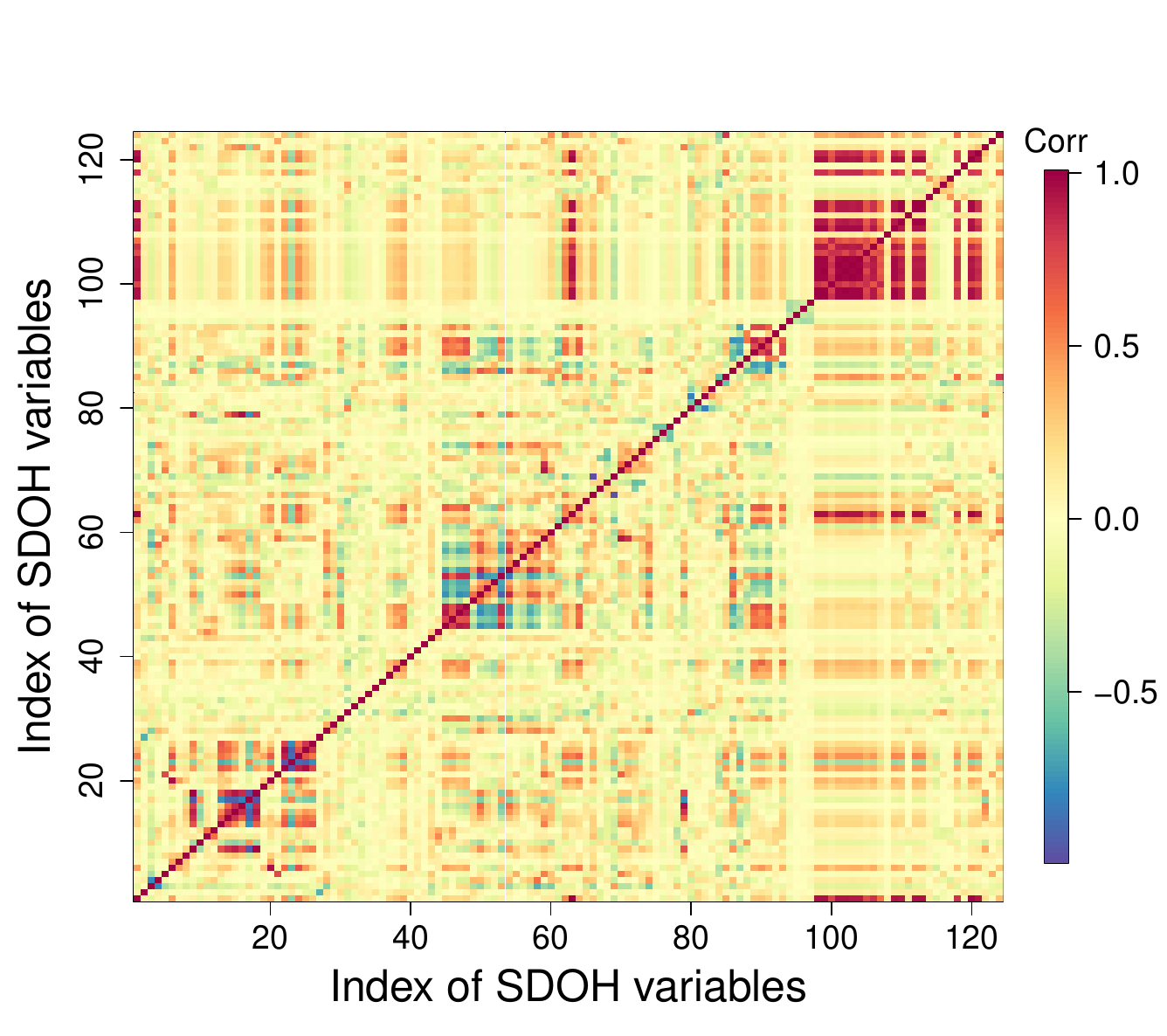}
		\caption{The correlation matrix for the SDOH variables (indexed by numbers). The correlation of the  longitudinal data between the $j$th SDOH and $j^\prime$th SDOH is calculated as $C_{j,j^\prime}/\sqrt{C_{j,j}C_{j^\prime,j^\prime}}$, where $C_{j,j^\prime}=
			n^{-1} \sum_{i=1}^nS_i^{-1}\sum_{s=1}^{S_i}\big\{X_{ij}(t_{is})-\hat{\mu}_{j}(t_{is})\big\}\big\{X_{ij^\prime}(t_{is})-\hat{\mu}_{j^\prime}(t_{is})\big\}$, with $\hat{\mu}_{j}(\cdot)$ being the estimated mean function for the $j$th county. Here, $X_{ij}(t_{is})$ is the observed data of the $j$th SDOH at the $i$th county on year $t_{is}$, for $i=1,\dots, n$, $j=1,\dots,  p$, and $s=1,\dots,  S_i$. The definition of $X_{ij}(t_{is})$, $n$, $p$, and $S_i$ can be found in Section \ref{sec: flr}.}
		\label{fig: sdoh cor}
	\end{figure}

	In 2020, the Agency for Healthcare Research and Quality (AHRQ) released
	a SDOH database to enhance understanding of community-level factors, healthcare
	quality and delivery, and individual health. The SDOH database contains
	yearly records of $345$ SDOH collected from 3226 counties from 2009 to
	2018. These SDOH are classified into five domains: (1) social context,
	such as age, race and ethnicity, and veteran status, (2) economic context,
	such as income and unemployment, (3) education, (4) physical infrastructure,
	such as housing, food insecurity, and transportation, and (5) health care
	contexts, such as health insurance coverage and health care access. To
	study the association between SDOH and stroke mortality, we connect the
	SDOH database with a stroke mortality data provided by the Interactive
	Atlas of Heart Disease and Stroke at the CDC on the county level. The stroke
	mortality database was originally compiled from two data sources: (1) the
	National Vital Statistics System at the National Center for Health Statistics
	and (2) the hospital discharge data from the Centers for Medicare \& Medicaid
	Services' Medicare Provider Analysis and Review (MEDPAR) file. All data
	and materials used in this analysis are publicly available at the AHRQ
	website: \url{https://www.ahrq.gov/sdoh/index.html} and CDC website
	\url{https://www.cdc.gov/dhdsp/maps/atlas/index.htm}. Since the AHRQ database
	is HIPAA (Health Insurance Portability and Accountability Act) compliant,
	our data do not require to be reviewed by an institutional review board.

	It is worth noting that the SDOH dataset contains missing values. Following
	the approach of a previous study on the dataset (\citet{son2023social}),
	we exclude the SDOH variables that have a missing proportion of more than
	60\%. The average missing proportion of the remaining variables is 3.6\%.
	For these missing values, we employ a K-Nearest Neighbors (KNN) method
	(\citet{kowarik2016imputation}) to impute the SDOH data, ensuring that the
	SDOH and stroke mortality data are aligned in time for each county. The
	detailed implementation of the KNN is provided in Part A of the Supplementary
	Material.

	\begin{figure}[ht]
		\centering
		\includegraphics[width=\linewidth]{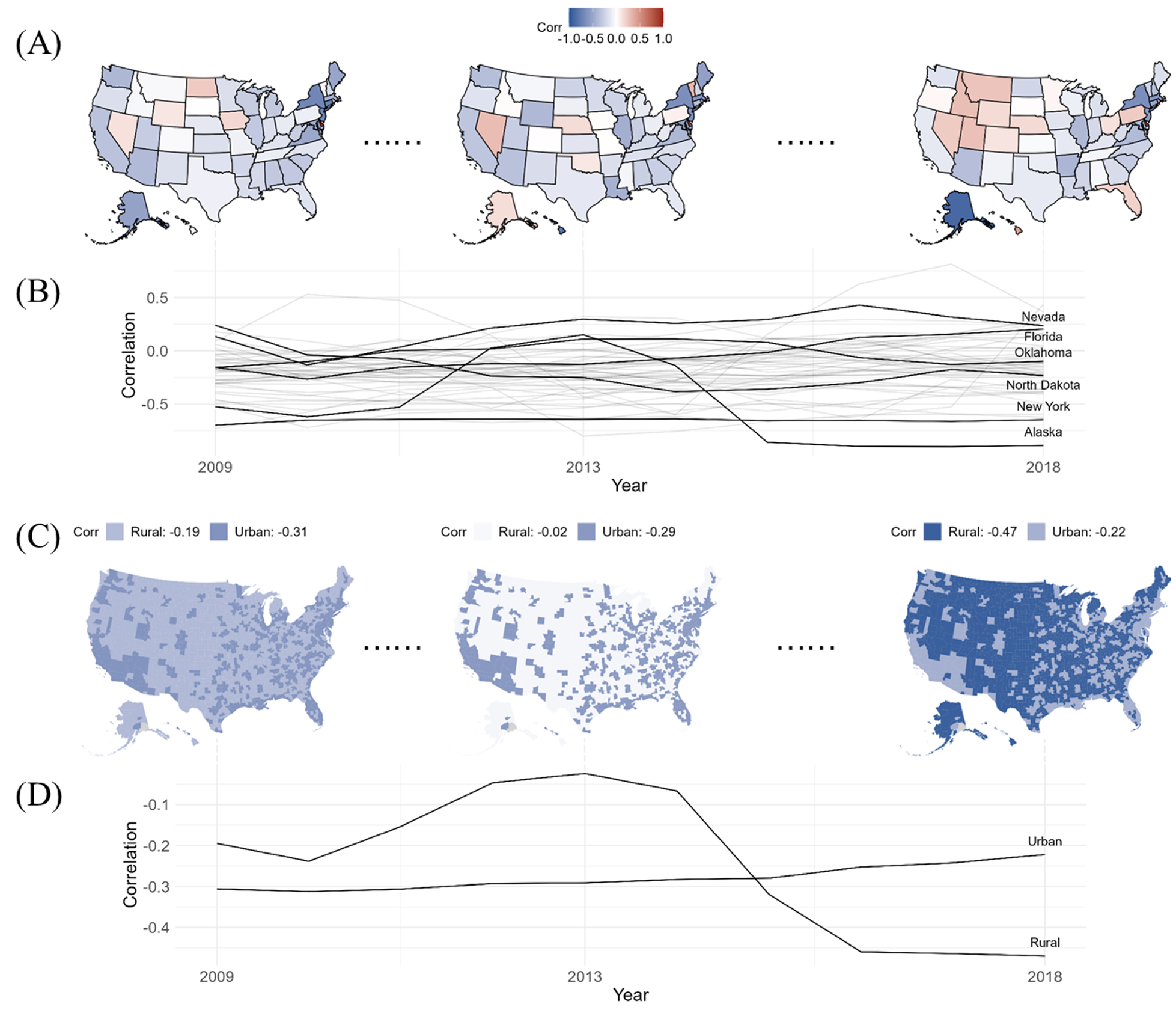}
		\caption{Statewise and rural-urban Pearson correlations between the stroke
			mortality and PAPL. Panels (A) and (C) are the geographical maps of correlations
			for the years 2009, 2013, and 2018, respectively, for two types of division
			strategies. Panels (B) and (D) are the statewise and rural-urban longitudinal
			correlations from 2009--2018, respectively.}
		\label{fig: cormaplong}
	\end{figure}

	As mentioned in the \hyperref[sec: intro]{Introduction}, significant collinearity may exist among
	the SDOH data. This phenomenon is observed in our dataset, where the longitudinal
	data between some of the SDOH exhibit significant correlations, as illustrated
	in Figure~\ref{fig: sdoh cor}. Besides, coinciding with the previous studies
	(\citet{villablanca2016outcomes,he2021trends,son2023social}), evident regional
	disparities and longitudinal variations can also be observed in the SDOH-stroke
	mortality associations from our dataset. To demonstrate this, we calculate
	the yearly Pearson correlation between stroke mortality and the percentage
	of Asian and Pacific Island language speakers (PAPL), an important sociocultural
	factor in the SDOH dataset. To calculate the yearly correlations, we adopt
	difference strategies to divide counties in the U.S. into several regions,
	followed by calculating the correlations in each divided region. Specifically,
	we illustrate the correlations based on two types of divisions: statewise
	division (\citet{zelko2023geographically}) and rural-urban division
	(\citet{son2023social}). The yearly correlations from 2009--2018 for each
	state, or for rural and urban areas, are presented in panels (B) and (D)
	of Figure~\ref{fig: cormaplong}, respectively. Besides, we also present
	the geographical maps of correlations for the years 2009, 2013, and 2018
	in panels (A) and (C) of Figure~\ref{fig: cormaplong}, respectively, for
	two types of division strategies.

	From the maps in Figure~\ref{fig: cormaplong}, we observe significant regional
	disparities in the correlations among different states as well as between
	rural and urban areas. These correlations all exhibit changes over time.
	In addition, we find that the longitudinal correlations in rural and urban
	areas are quite different from the correlations from the states, as displayed
	in panels (B) and (D) of Figure~\ref{fig: cormaplong}.  These results
	suggest that the outcomes of longitudinal associations are sensitive to
	the region division strategy, and different region divisions may lead to
	inconsistent conclusions for the stroke management. To address this issue,
	we require a reasonable data-driven method for the region division in exploring
	SDOH-stroke mortality associations.

	\subsection{Finite Mixture of Functional Linear Concurrent Models}
	\label{sec: flr}
	
	For $i=1,\ldots,n$ and $t\in \mathcal{T}$, let
	$Y_{i}(t)\in \mathbb{R}$ and
	$\boldsymbol{X}_{i}(t):=(X_{i1}(t),\dots,X_{ip}(t))^{\top}\in
	\mathbb{R}^{p}$ represent the stroke mortality and SDOH data in the
	$i$th county at time $t$, respectively, where $n$ is the number of counties
	and $\mathcal{T}$ is the observed time period. Here $Y_{i}(\cdot )$ and
	$\boldsymbol{X}_{i}(\cdot )$ are considered the functional response and
	functional covariate for the $i$th subject, respectively, with $n$ and
	$p$ being the sample size and dimension of the covariates. Without loss
	of generality, we assume that $\mathcal{T}=[0,1]$ in this article.

	We focus on the relation between $Y_{i}(t)$ and
	$\boldsymbol{X}_{i}(t)$ for all $t\in [0,1]$, referred to as longitudinal
	associations in what follows. To cluster the longitudinal associations
	of different counties, we assume the samples
	$\{Y_{i}(\cdot ),\boldsymbol{X}_{i}(\cdot );i=1,\ldots,n\}$ can be divided
	into $K$ clusters, following a finite mixture of functional linear concurrent
	models (FMFLCM):
	%
	\begin{equation}
		Y_{i}(t)=\sum_{j=1}^{p}X_{ij}(t)
		\beta _{jk}(t)+\varepsilon _{ik}(t),\quad t\in [0,1] \text{ if }
		Z_{i}=k, \label{eq: model} 
	\end{equation}
	where $Z_{i}$ is the cluster membership for the $i$th subject,
	$\beta _{jk}(\cdot )$ is the functional coefficient for the $k$th group
	reflecting the longitudinal association, and
	$\{\{\varepsilon _{ik}(t);t\in [0,1]\};i=1,\ldots,n, k=1,\ldots,K
	\}$ are mean-zero Gaussian processes on $[0,1]$ that are independent across
	different $i$ and $k$. We assume that $Z_{1},\ldots,Z_{n}$ are independent
	and identically distributed (i.i.d.) samples from a multinomial distribution
	on ${1,\ldots,K}$ with $\mathbb{P}(Z_{i}=k)=\pi _{k}$ and
	$\operatorname{var}(\varepsilon _{ik}(t))=\sigma _{k}^{2}$, $i=1,\ldots,n$ and
	$t\in [0,1]$, where the error process
	$\{\varepsilon _{ik}(t);t\in [0,1]\}$ may possess temporal correlations
	among different time $t$. By taking a fixed time $t\in \mathcal{T}$ for
	the functional objects in \eqref{eq: model}, this model is essentially
	a finite mixture of linear regression models
	(\citet{peel2000finite, khalili2007variable,khalili2013regularization}),
	capturing heterogeneity in the associations between
	$\{Y_{i}(t);i=1,\ldots,n\}$ and
	$\{(X_{i1}(t),\ldots,X_{ip}(t))^{\top};i=1,\ldots,n\}$ for each
	$t$. We generalize the mixture of linear regression models to its functional
	version in \eqref{eq: model}, aiming to account for the heterogeneity and
	smoothness of $\beta _{jk}(\cdot )$ during model estimation.

	In the following, let
	$\boldsymbol{\beta }(t)=  \{\beta _{jk}(t);j=1,\ldots,p;k=1,
	\ldots,K  \}$,
	$\boldsymbol{\sigma }^{2}:=(\sigma _{1}^{2},\ldots,\sigma _{K}^{2})^{
		\top}$, and $\boldsymbol{\pi }:=(\pi _{1},\ldots,\pi _{K})^{\top}$. We
	use $\boldsymbol{\beta }_{\cdot k}(t)$ and
	$\boldsymbol{\beta }_{j\cdot}(t)$ to denote the $k$th column and the
	$j$th row of $\boldsymbol{\beta }(t)$, respectively. For ease of notation,
	we might use $\boldsymbol{\beta }$, $\boldsymbol{\beta }_{\cdot k}$,
	$\boldsymbol{\beta }_{j\cdot}$, and $\beta _{jk}$ to represent
	$\boldsymbol{\beta }(\cdot )$,
	$\boldsymbol{\beta }_{\cdot k}(\cdot )$,
	$\boldsymbol{\beta }_{j\cdot}(\cdot )$, and $\beta _{jk}(\cdot )$, respectively.

	We assume the functional variables $Y_{i}(\cdot )$ and	$\boldsymbol{X}_{i}(\cdot )$ in FMFLCM \eqref{eq: model} are only observed on a finite time grid, denoted by	$ \{ (Y_{i}(t_{is}),\boldsymbol{X}_{i}(t_{is}) )$;
	$i=1,\ldots,n, s=1,\ldots,S_{i} \}$, where $S_{i}$ is the observed	number of time points for subject $i$, and
	$\{t_{is};i=1,\ldots,n, s=1,\ldots,S_{i}\}$ are time points contained in $[0,1]$ that can be varying across different subjects. Based on the above notations, we estimate the parameters $\boldsymbol{\Phi }:=(\boldsymbol{\beta },\boldsymbol{\sigma }^{2},\boldsymbol{\pi })$ by maximizing the composite log-likelihood of the data	(\citet{varin2011overview})
	\begin{align}
		l(\boldsymbol{\Phi }) &=\sum_{i=1}^{n}\sum
		_{s=1}^{S_{i}}\operatorname{log} \Biggl\{\sum
		_{k=1}^{K}\pi _{k} f \bigl(Y_{i}(t_{is});
		\boldsymbol{X}_{i}(t_{is})^{
			\top}\boldsymbol{\beta
		}_{\cdot k}(t_{is}),\sigma _{k}^{2} \bigr)
		\Biggr\}, \label{eq: likelihood1} 
	\end{align}
	where $f(\cdot;\mu,\sigma ^{2})$ is a Gaussian density function with
	mean $\mu $ and variance $\sigma ^{2}$. In the composite likelihood, we
	do not consider the temporal dependencies within residual processes
	$\varepsilon _{ik}(\cdot )$ in order to reduce computational costs required
	for model estimation; similar strategies have been adopted in the estimation
	of functional regressions
	(\citet{huang2002varying,zhu2012multivariate}).

	In general, optimizing $\boldsymbol{\beta }$ in	\eqref{eq: likelihood1} is an ill-posed problem due to the infinite-dimensional nature of $\boldsymbol{\beta }$. Furthermore, estimating $\boldsymbol{\beta }$ may suffer from the curse of dimensionality when $p$ is relatively large. In such cases the estimated functional coefficients can be far from their true values or may not even exist (\citet{wang2008variable, yi2015regularized}). To address these issues, we introduce a class of penalties in Section~\ref{sec: penalty} to regularize	the smoothness and sparsity of $\boldsymbol{\beta }$ simultaneously. Based
	on these penalties, we develop a regularized expectation maximization algorithm	in Section~\ref{sec: EM} to simultaneously perform clustering of subjects, estimation of $\boldsymbol{\beta }$, and variable selection of the covariates.

	\subsection{Sparsity and Smoothness Pursued Penalties}
	\label{sec: penalty}

	In this subsection we propose a class of smoothly clipped absolute deviation (SCAD) type penalties (\citet{fan2001variable}) to adjust the aforementioned ill-posed problems. The SCAD penalty takes the form
	\begin{equation*}
		\boldsymbol{P}_{\text{SCAD}} ( u; \lambda ) = %
		\begin{cases} \lambda u
			& \text{if } 0 \leq u \leq \lambda,
			\\
			-\frac {u^{2} - 2\gamma \lambda u + \lambda ^{2}}{2(\gamma - 1)} & \text{if } \lambda < u < \gamma \lambda,
			\\
			\frac {(\gamma + 1)\lambda ^{2}}{2} & \text{if } u \geq \gamma \lambda, \end{cases} %
	\end{equation*}
	where $u$ is a scalar parameter to be penalized, $\lambda $ is a tuning	parameter, and $\gamma $ is a hyperparameter chosen to be 3.7, as suggested in \cite{fan2001variable}. The SCAD penalty has been shown to possess desirable theoretical properties for variable selection in various regression models
	(\citet{fan2001variable,wang2007group,wang2008variable}). Here we extend the penalty to the functional objects $\boldsymbol{\beta }$, leading to the functional SCAD (FS) penalty
	%
	\begin{equation}
		\boldsymbol{P}_{\text{FS}}(\boldsymbol{\beta };{\lambda _{1},
			\ldots, \lambda _{K}},r):=\sum_{j=1}^{p}
		\sum_{k=1}^{K}\boldsymbol{P}_{
			\text{SCAD} }
		\bigl( \lVert \beta _{jk} \rVert _{r};{\lambda
			_{k}}\bigr), \label{eq: hybrid norm} 
	\end{equation}
	where
	$\Vert \beta _{jk}\Vert _{r}=\sqrt{\Vert \beta _{jk}\Vert ^{2}+r \Vert \beta _{jk}^{(2)}\Vert ^{2}}$, with	$\Vert f\Vert =\sqrt{\int _{0}^{1} f^{2}(t)\, \mathrm{d}t}$ being the	$L_{2}$ norm for a function $f$. The norm	$\Vert \beta _{jk}\Vert _{r}$ simultaneously measures the magnitude and smoothness of $\beta _{jk}$ (\citet{meier2009high}), in which $r$ leverages	the two terms in the norm. By penalizing $\boldsymbol{\beta }$ using \eqref{eq: hybrid norm}, we can regularize the smoothness of each $\beta _{jk}$ to deal with its functional structure, and shrink the $\beta _{jk}$ with small norm $\Vert \beta _{jk}\Vert _{r}$ to zero
	(\citet{huang2009analysis}) for the purpose of variable selection.

	Recall that our SDOH functional covariates exhibit significant collinearity, as illustrated in Figure~\ref{fig: sdoh cor}. This collinearity may increase
	variance and potentially lead to misspecification in variable selection
	(\citet{zou2005regularization}). To address this issue, we adopt a similar
	strategy as the elastic net (\citet{zou2009adaptive}) to add an
	$L_{2}$ term to the FS penalty \eqref{eq: hybrid norm}. As such, we obtain
	the functional SCAD-Net (FS-Net) penalty
	%
	\begin{equation}
		\boldsymbol{P}_{\text{FS-Net}}(\boldsymbol{\beta };\lambda _{1},
		\ldots,\lambda _{K},\rho,r):=\sum_{j=1}^{p}
		\sum_{k=1}^{K} \bigl\{ \rho
		\boldsymbol{P}_{\text{SCAD}} \bigl( \lVert \beta _{jk} \rVert
		_{r}; \lambda _{k} \bigr)+(1-\rho ) \lambda _{k}
		\lVert \beta _{jk} \rVert _{r}^{2} \bigr\},
		\label{eq: penalty FSCADL2} 
	\end{equation}
	where $\rho \in [0,1]$ is a tuning parameter controlling the proportions
	between the FS penalty and the $L_{2}$ term. The above penalty not only
	maintains the properties of the FS penalty but also reduces variance and
	addresses misspecification in variable selections by penalizing the magnitude
	of
	$\sum_{j=1}^{p}\sum_{k=1}^{K}\lambda _{k}\Vert \beta _{jk}\Vert _{r}^{2}$.

	In general, variable selection with FS or FS-Net does not guarantee that the selected covariates from different clusters are the same. This may not be suitable for some applications where identifying important variables for all clusters is necessary. In our case, the purpose of variable selection	for SDOH is to aid clustering based on the selected covariates. Therefore,	we require that the selected SDOH for all clusters be the same within estimation procedures. To facilitate this kind of investigation, we propose shrinking
	${\beta _{jk}; k=1,\ldots, K}$ to zero functions simultaneously, referred to as cluster-invariant sparsity for variable selection. For this we modify	the FS-Net penalty into the functional group SCAD-Net (FGS-Net) penalty
	%
	\begin{equation}
		\boldsymbol{P}_{\text{FGS-Net}}(\boldsymbol{\beta };\lambda,\rho,r):= \sum
		_{j=1}^{p} \bigl\{\rho \boldsymbol{P}_{\text{SCAD} }
		\bigl( \lVert \boldsymbol{\beta }_{j\cdot} \rVert _{r};\lambda
		\bigr)+(1-\rho )\lambda \lVert \boldsymbol{\beta }_{j\cdot} \rVert
		_{r}^{2} \bigr\}, \label{eq: penalty FGSCADL2} 
	\end{equation}
	where
	$\Vert \boldsymbol{\beta }_{j\cdot}\Vert _{r}=\sqrt{\sum_{k=1}^{K}	\Vert \beta _{jk}\Vert _{r}^{2}}$. By grouping $  \{\beta _{kj};k=1,\ldots, K  \}$ for each $j$ together in the norm, the FGS-Net not only yields cluster-invariant sparsity but also	improves statistical efficiency for variable selection by borrowing strength
	across all clusters.

	\subsection{Regularized EM Algorithm}
	\label{sec: EM}

	In this subsection we apply an EM algorithm to proceed with the parameter estimation of FMFLCM in \eqref{eq: model}. We first introduce a latent	variable $z_{ik}:=I(Z_{i}=k)$ to represent the cluster membership of the subject $i$, where $I(\cdot )$ is an indicator function. Denote $\boldsymbol{Z}$ as $\{z_{ik};i=1,\ldots,n,k=1,\ldots,K\}$. The composite log-likelihood of FMFLCM containing $\boldsymbol{Z}$ is then given by
	%
	\begin{align}
		l_{C}(\boldsymbol{\Phi },\boldsymbol{Z})&=\sum
		_{i=1}^{n}\sum_{k=1}^{K}z_{ik}
		\sum_{s=1}^{S_{i}} \bigl[\operatorname{log}\pi
		_{k}+\operatorname{log} \bigl\{f \bigl(Y_{i}(t_{is});
		\bigl\{\boldsymbol{X}_{i}(t_{is})\bigr\}^{\top}
		\boldsymbol{\beta }_{\cdot k}(t_{is}), \sigma _{k}^{2}
		\bigr) \bigr\} \bigr]. \label{eq: complete lik} 
	\end{align}
	To address the ill-posed issue for estimating functional coefficients
	$\boldsymbol{\beta }$, we develop a Regularized EM (REM) algorithm by incorporating	penalty functions proposed in Section~\ref{sec: penalty}. In the following	we demonstrate the E-step and M-step of the REM algorithm equipped with the FGS-Net penalty \eqref{eq: penalty FGSCADL2}, while those equipped with FS penalty \eqref{eq: hybrid norm} or FS-Net penalty \eqref{eq: penalty FSCADL2} can be obtained similarly.
	
	\subsubsection{Procedures of REM algorithm}
		\label{sec: Estep}
	Let $\boldsymbol{\Phi }^{(m-1)}:=(\boldsymbol{\beta }^{(m-1)},(\boldsymbol{\sigma }^{2})^{(m-1)},\boldsymbol{\pi }^{(m-1)})$ be the parameters updated at the $(m-1)$th iteration. The REM algorithm iterates between	the following E-step and M-step:
	
	\textbf{E-step}: With the parameter $\boldsymbol{\Phi }^{(m-1)}$, we first compute the conditional expectation of $z_{ik}$, given the samples	$\boldsymbol{Y}=\{Y_{i}(t_{is});i=1,\ldots,n,s=1,\ldots.S_{i}\}$ and	$\boldsymbol{X}=\{\boldsymbol{X}_{i}(t_{is});i=1,\ldots,n,s=1,	\ldots,S_{i}\}$,
	%
	\begin{equation}\begin{split}
			\omega _{ik}^{(m)}&:=\mathbb{E}_{\boldsymbol{\Phi }^{(m-1)}}(z_{ik}
			| \boldsymbol{Y},\boldsymbol{X})\\
			&= \frac{{\pi _{k}^{(m-1)}}\sum_{s=1}^{S_{i}}f  (Y_{i}(t_{is}),\boldsymbol{X}_{i}(t_{is})^{\top}\boldsymbol{\beta }_{\cdot k}^{(m-1)}(t_{is}),(\sigma _{k}^{2})^{(m-1)}  )}{
				\sum_{k=1}^{K}{\pi _{k}^{(m-1)}}\sum_{s=1}^{S_{i}}f  (Y_{i}(t_{is}),\boldsymbol{X}_{i}(t_{is})^{\top}\boldsymbol{\beta }_{\cdot k}^{(m-1)}(t_{is}),(\sigma _{k}^{2})^{(m-1)}  )}. \label{eq: omega update} 
	\end{split}\end{equation}
	After that, we calculate the expectation of the log-likelihood in	\eqref{eq: complete lik} conditioning on $\boldsymbol{Y}$ and	$\boldsymbol{X}$, given the parameter $\boldsymbol{\Phi }^{(m-1)}$. This
	leads to the Q function
	%
	\begin{equation}
		Q\bigl(\boldsymbol{\Phi } | \boldsymbol{\Phi }^{(m-1)}\bigr):=\sum
		_{i=1}^{n} \sum_{k=1}^{K}
		\omega _{ik}^{(m)}\sum_{s=1}^{S_{i}}
		\bigl[\operatorname{log} \pi _{k}+\operatorname{log} \bigl\{f \bigl(Y_{i}(t_{is}),
		\bigl\{\boldsymbol{X}_{i}(t_{is}) \bigr\}^{\top}
		\boldsymbol{\beta }_{\cdot k}(t_{is}),\sigma _{k}^{2}
		\bigr) \bigr\} \bigr]. \label{eq: Qfunction} 
	\end{equation}

	\textbf{M-step}: To facilitate the update of $\boldsymbol{\beta }$, we incorporate	the FGS-Net penalty \eqref{eq: penalty FGSCADL2} into the Q function	\eqref{eq: Qfunction},
	%
	\begin{align}
		&Q^{\text{pen}} \bigl(\boldsymbol{\Phi } | \boldsymbol{\Phi }^{(m-1)};
		\lambda,\rho,r \bigr):= Q \bigl(\boldsymbol{\Phi } | \boldsymbol{\Phi
		}^{(m-1)} \bigr)- \boldsymbol{P}_{\text{FGS-Net}} (\boldsymbol{\beta };
		\lambda,\rho,r ). \label{eq: pen Q} 
	\end{align}
	According to \eqref{eq: pen Q}, we separately update the parameters	$\boldsymbol{\beta }$, $\boldsymbol{\sigma }^{2}$, and	$\boldsymbol{\pi }$, given the current tuning parameters	$\lambda,\rho,r$. In detail, holding $\boldsymbol{\sigma }^{2}$ and	$\boldsymbol{\pi }$ fixed at their previous values at the $(m-1)$th iteration,	we update $\boldsymbol{\beta }$ as
	%
	\begin{align}
		\boldsymbol{\beta }^{(m)}=\operatorname{argmax}_{\boldsymbol{\beta }}{Q}^{
			\text{pen}}
		\bigl( \bigl(\boldsymbol{\beta },\bigl(\boldsymbol{\sigma }^{2}
		\bigr)^{(m-1)}, \boldsymbol{\pi }^{(m-1)} \bigr) | \boldsymbol{\Phi
		}^{(m-1)}; \lambda,\rho,r \bigr). \label{eq: beta update} 
	\end{align}
	Once obtaining $\boldsymbol{\beta }^{(m)}$, we update
	$\boldsymbol{\pi }^{(m)}$ by
	%
	\begin{equation}
		\pi _{k}^{(m)}=\frac{\sum_{n=1}^{N}\omega _{ik}^{(m)}}{N},\quad  k=1, \ldots,K
		\label{eq: pi update} 
	\end{equation}
	and update $(\boldsymbol{\sigma }^{2})^{(m)}$ by
	%
	\begin{equation}
		\bigl(\sigma _{k}^{2}\bigr)^{(m)}=
		\frac{
			\sum_{i=1}^{N}\omega _{ik}^{(m)}\sum_{s=1}^{S_{i}}   \{Y_{i}(t_{is})-\boldsymbol{X}_{i}(t_{is})^{\top}\boldsymbol{\beta }_{\cdot k}^{(m)}(t_{is})  \}^{2}
		}{\sum_{i=1}^{N}S_{i}\omega _{ik}^{(m)}},\quad k=1,\ldots,K, \label{eq:sigma update} 
	\end{equation}
	where $\omega _{ik}^{(m)}$ is defined in \eqref{eq: omega update}. Given	\eqref{eq: beta update}, \eqref{eq: pi update}, and	\eqref{eq:sigma update}, we subsequently update	$\boldsymbol{\Phi }^{(m)}=(\boldsymbol{\beta }^{(m)},(\boldsymbol{\sigma }^{2})^{(m)},\boldsymbol{\pi }^{(m)})$ until the convergence.	The stopping criteria of the REM algorithm are provided in Section B of
	the Supplementary Material.
	
	Note that the main efforts for the above procedures lie in the optimization
	in \eqref{eq: beta update}. We demonstrate this process in detail in Sections~\ref{sec: Optimization of FGSCAD-Net Regularization}.

	\subsubsection{Optimization of FGS-Net Regularization} \label{sec: Optimization of FGSCAD-Net Regularization}
	
	Since we penalize the smoothness of $\boldsymbol{\beta }$ in \eqref{eq: penalty FGSCADL2} via controlling their second-order derivatives,	we adopt a similar strategy of smoothing spline	(\citet{perperoglou2019review}) to represent the functional coefficients.	Specifically, we parameterize $\beta _{jk}(t)$ by the cubic spline basis	functions $\psi _{1}(t),\ldots,\psi _{L}(t)$, with the knots of spline
	functions given as $\bigcup_{i=1}^{n}\{t_{is};s=1,\ldots,S_{i}\}$. As such,	$\beta _{jk}(t)$ is represented by
	%
	\begin{equation}
		\beta _{jk}(t)=\boldsymbol{b}_{jk}^{\top}\boldsymbol{
			\Psi }(t), \quad t \in [0,1], \label{eq: betabpsi} 
	\end{equation}
	where	$\boldsymbol{b}_{jk}=(b_{jk1},\ldots,b_{jkL})^{\top}\in \mathbb{R}^{L}$	and	$\boldsymbol{\Psi }(t)=(\psi _{1}(t),\ldots,\psi _{L}(t))^{\top}\in	\mathbb{R}^{L}$. After that, we maximize \eqref{eq: beta update} by substituting	\eqref{eq: betabpsi} into the Q function \eqref{eq: pen Q}. We use
	$\boldsymbol{D}^{\top}\boldsymbol{D}$ to denote the Cholesky decomposition
	of the nonnegative definite matrix
	\begin{equation*}
		\int _{0}^{1} \boldsymbol{\Psi }(t) \bigl\{
		\boldsymbol{\Psi }(t)\bigr\}^{\top}+r \boldsymbol{\Psi }^{\prime \prime }(t)
		\bigl\{\boldsymbol{\Psi }^{\prime \prime }(t)\bigr\}^{\top} \,\mathrm{d}t,
	\end{equation*}
	where $\boldsymbol{D}\in \mathbb{R}^{L\times L}$ is an upper triangular	matrix. Given this, define	$\boldsymbol{\alpha }_{jk}=\boldsymbol{D}\boldsymbol{b}_{jk}\in	\mathbb{R}^{L}$ and	$\boldsymbol{\alpha }_{j\cdot}= (\boldsymbol{\alpha }_{j1}^{\top},	\ldots,\boldsymbol{\alpha }_{jK}^{\top} )^{\top}\in \mathbb{R}^{LK}$.	We rewrite $\Vert \boldsymbol{\beta }_{j\cdot}\Vert _{r}$ as
	\begin{align*}
		\lVert \boldsymbol{\beta }_{j\cdot} \rVert _{r}&=\sqrt{
			\lVert \boldsymbol{\beta }_{j\cdot} \rVert ^{2}+r \bigl\lVert (
			\boldsymbol{\beta }_{j
				\cdot})^{\prime \prime } \bigr\rVert ^{2}}
		\nonumber
		\\
		&=\sqrt{\sum_{k=1}^{K}
			\int _{0}^{1}\boldsymbol{b}_{jk}^{\top}
			\boldsymbol{\Psi }(t) \bigl\{\boldsymbol{\Psi }(t)\bigr\}^{\top}
			\boldsymbol{b}_{jk}+r \boldsymbol{b}_{jk}^{\top}
			\boldsymbol{\Psi }^{\prime \prime }(t) \bigl\{ \boldsymbol{\Psi }^{\prime \prime }(t)
			\bigr\}^{\top}\boldsymbol{b}_{jk} \,\mathrm{d}t}
		\nonumber
		\\
		&=\sqrt{\sum_{k=1}^{K}(
			\boldsymbol{D}\boldsymbol{b}_{jk})^{\top} \boldsymbol{D}
			\boldsymbol{b}_{jk}}
		= \lVert \boldsymbol{\alpha }_{j\cdot} \rVert ,
	\end{align*}
	where we abuse the notation $\Vert \cdot \Vert $ to denote the Euclidean norm of	a vector. We further denote	$ \boldsymbol{\alpha }=  (\boldsymbol{\alpha }_{1\cdot}^{\top},	\ldots, \boldsymbol{\alpha }_{p\cdot}^{\top} )^{\top} \in	\mathbb{R}^{pLK} \text{ and } \boldsymbol{h}_{ij}(t) = X_{ij}(t)	\cdot   (   ( \psi _{1}(t), \ldots, \psi _{L}(t)   )	\boldsymbol{D}^{-1}   )^{\top} \in \mathbb{R}^{L} $. With these notations	we can transform the optimization \eqref{eq: beta update} into a standard	group SCAD-$L_{2}$ optimization problem (\citet{zeng2014group}) as follows:
	%
	\begin{align}
		& \operatorname{argmax}_{\boldsymbol{\alpha }} \Biggl[L_{m}(\boldsymbol{\alpha }) -		\sum_{j=1}^{p} \bigl\{\rho		\boldsymbol{P}_{\text{SCAD}} \bigl( \lVert \boldsymbol{\alpha }_{j\cdot}		\rVert ;\lambda \bigr)+(1-\rho ) \lambda \lVert \boldsymbol{\alpha		}_{j\cdot} \rVert ^{2} \bigr\} \Biggr], \label{eq: alpha tilde optim}
	\end{align}
	where
	\begin{equation*}
		L_{m}(\boldsymbol{\alpha }) = \sum_{i=1}^{n}		\sum_{k=1}^{K} \omega _{ik}^{(m)}	\sum_{s=1}^{S_{i}} \Biggl[ \log \pi		_{k}^{(m-1)} + \log \Biggl\{ f \Biggl( Y_{i}(t_{is}),		\sum_{j=1}^{p} \boldsymbol{h}^{\top}_{ij}(t_{is})		\boldsymbol{\alpha }_{jk}, \bigl(\sigma _{k}^{2}		\bigr)^{(m-1)} \Biggr) \Biggr\} \Biggr].
	\end{equation*}
	The above optimization can be efficiently solved by the group coordinate	descent algorithm (\citet{breheny2015group}). Once we obtain the maximizer	of $\boldsymbol{\alpha }_{jk}$ from \eqref{eq: alpha tilde optim}, denoted	by ${\boldsymbol{\alpha }}^{(m)}_{jk}$, we can estimate the $(j,k)$th element	of ${\boldsymbol{\beta }}^{(m)}$ by
	%
	\begin{equation}
		\beta _{jk}^{(m)}(t) = \bigl( \boldsymbol{D}^{-1}
		\boldsymbol{\alpha}_{jk}^{(m)} \bigr)^{\top}
		\boldsymbol{\Psi }(t),\quad  t \in [0,1]. \label{eq: alpha to beta} 
	\end{equation}
	For each $j$, the penalty in \eqref{eq: alpha tilde optim} encourages	${\boldsymbol{\alpha }}_{jk}^{(m)}$, $k=1,\ldots,K$, to be zero vectors	simultaneously. The functional coefficients	$\{{\beta}_{jk}^{(m)};k=1,\ldots,K\}$ are smooth functions, and they become	zero functions simultaneously as $\lambda $ increases, satisfying the cluster-invariant	sparsity.
	
	We similarly establish the REM algorithms for the FS penalty
	\eqref{eq: hybrid norm} or FS-Net penalty \eqref{eq: penalty FSCADL2} by
	modifying the penalty function in \eqref{eq: alpha tilde optim} as
	\begin{align}
		&\text{FS:}\quad\sum_{k=1}^{K}\sum		_{j=1}^{p} \boldsymbol{P}_{			\text{SCAD}} \bigl(
		\lVert \boldsymbol{\alpha }_{jk} \rVert ;\lambda _{k}		\bigr),		\nonumber		\\
		&\text{FS-Net:}\quad\sum_{k=1}^{K}\sum		_{j=1}^{p} \bigl\{\rho \boldsymbol{P}_{\text{SCAD}}		\bigl( \lVert \boldsymbol{\alpha }_{jk} \rVert ;\lambda_{k} \bigr)+(1-\rho )\lambda _{k} \lVert \boldsymbol{
			\alpha }_{jk} \rVert ^{2} \bigr\}.
		\nonumber
	\end{align}
	In these two cases, $\{{\boldsymbol{\alpha }}_{jk}^{(m)}$,	$k=1,\ldots,K\}$ are not encouraged to be zero vectors simultaneously,	only obtaining the cluster-variant sparsity during estimation.

	\subsection{Tuning Strategies}
	\label{sec: tuning}

	The REM algorithm requires tuning of three hyperparameters	$\lambda $, $\rho $, and $r$. The traditional strategy for choosing	$\lambda $, $\rho $, and $r$ needs to run the entire algorithm for all	candidate combinations of $(\lambda, \rho, r)$. This is computationally	inefficient since there are numerous choices of candidate combinations	that need to be examined. Inspired by the path-fitting algorithm	(\citet{breheny2015group}) for a fast tunning of $\lambda $, we modify the	traditional tuning strategy for our REM algorithm. Instead of running a	complete REM for each candidate of $\lambda $, we propose to tune	$\lambda $ in each M-step of the REM algorithm; refer to Part B.1 in the	Supplementary Material for the implementation details. Using this approach,	we can accelerate the tuning process by employing a path-fitting algorithm	to select $\lambda $, similar to the existing literature	(\citet{huang2009analysis,li2016supervised,cai2019chime}).
	
	For the hyperparameters $\rho $ and $r$, we run an entire REM for each	of their candidate combinations, nested with the aforementioned strategy	for tuning $\lambda $. Since $\rho $ and $r$ are imposed to penalize functional	smoothness and improve prediction performance, we adopt the Akaike Information
	Criterion (AIC) for the selection of $\rho $ and $r$, as AIC offers a predictive	criterion for model selection (\citet{ding2018model}). Specifically, the	AIC is calculated as
	%
	\begin{equation}
		\text{AIC}(\rho,r)=-2l\bigl(\widehat{\boldsymbol{\Phi }}(\rho,r)\bigr)+2
		\text{df}(\rho,r), \label{eq: aic rhor} 
	\end{equation}
	where $l(\cdot )$ is defined in \eqref{eq: likelihood1},	$\widehat{\boldsymbol{\Phi }}(\rho,r)$ is the converged parameters of	the REM algorithm, given the current choices of $\rho $ and $r$, and	$\text{df}(\rho,r)$ is the degree of freedom determined by $\rho $ and	$r$. In addition, we select the number of clusters $K$ using Bayesian information	criterion (BIC), a widely used criterion for model-based clustering procedure	(\citet{keribin2000consistent,james2003clustering,ding2018model,liang2021modeling}).
	The BIC is given as
	%
	\begin{align}
		\text{BIC}(K) &= -2l(\boldsymbol{\Phi }_{K}) + \text{df}(K) \cdot
		\log \Biggl(\sum_{i=1}^{n} S_{i}
		\Biggr), \label{eq:BIC} 
	\end{align}
	where $\boldsymbol{\Phi }_{K}$ denotes the converged parameter	$\widehat{\boldsymbol{\Phi }}(\rho,r)$ with $\rho $ and $r$ selected by	\eqref{eq: aic rhor} and the number of clusters being $K$, and	$\text{df}(K)$ is the degree of freedom determined by $K$. For the detailed	definitions of $\text{df}(\rho,r)$ and $\text{df}(K)$ in AIC	\eqref{eq: aic rhor} and BIC \eqref{eq:BIC}, please refer to
	\citet{breheny2015group}.

	\section{Simulation}
	\label{sec: simu}
	
	In this section, we conduct numerical simulations to assess the performances of the proposed method in Section~\ref{sec:method}, in comparison to other competing methods across three aspects: clustering, variable selection, and parameter estimation.

	\subsection{Data Generation}
	
	We generate the functional covariates $\boldsymbol{X}_{i}(\cdot )$,	$i=1,\ldots,n$, by	%
	\begin{equation*}
		\boldsymbol{X}_{i}(t)=\sum_{l=1}^{4}
		\boldsymbol{\theta }_{il}\psi _{l}(t),\quad \forall t\in [0,1],
		\label{eq:_xgen_VTEX1}
	\end{equation*}	%
	where $\psi _{1}(\cdot ),\ldots,\psi _{4}(\cdot )$ are the first four	nonconstant Fourier basis functions, and
	$\boldsymbol{\theta }_{il}\in \mathbb{R}^{p}$, for each $i$ and $l$, is	a random vector sampled from a mean-zero Gaussian distribution with the	covariance matrix	$(l^{-2}\alpha ^{|j-k|})_{1\leq j,k\leq p}\in \mathbb{R}^{p\times p}$.	The parameter $\alpha $ controls the dependence among covariates, with
	a higher value indicating stronger dependencies.
		
	To generate the functional coefficients attaining the cluster-invariant	sparsity, we set	$\boldsymbol{\beta }_{\cdot k}(t)\in \mathbb{R}^{p}$ as	%
	\begin{equation*}
		\boldsymbol{\beta }_{\cdot k}(t)=\bigl(f_{1k}(t),f_{1k}(t),f_{2k}(t),f_{2k}(t),f_{3k}(t),f_{3k}(t),0,
		\ldots,0\bigr)^{\top},\quad k=1,\ldots,K, t\in [0,1], \label{eq:_betagen_VTEX1}
	\end{equation*}	%
	where $f_{jk}(t)=f_{jk}^{*}(t)/ \Vert f_{jk}^{*} \Vert _{2}$, and	$f_{jk}^{*}(\cdot )$s are given by	%
	\[
\begin{aligned}
f_{11}^*(t) &= \sin\Bigl(\frac{\pi t}{2}+\frac{3\pi}{2}\Bigr)-t-\tfrac{1}{2}, \quad
& f_{12}^*(t) &= \big\{\cos(2\pi t)-1\big\}^2, \\[0.5em]
f_{13}^*(t) &= -f_{11}^*(t)+1, \quad
& f_{21}^*(t) &= \sin(2\pi t)-t+0.5, \\[0.5em]
f_{22}^*(t) &= \sin\Bigl(\frac{\pi t}{2}+\pi\Bigr), \quad
& f_{23}^*(t) &= -f_{21}^*(t)-0.5, \\[0.5em]
f_{31}^*(t) &= -\sin\Bigl(\frac{\pi t}{2}+\frac{3\pi}{2}\Bigr)-t-0.5, \quad
& f_{32}^*(t) &= -f_{12}^*(t), \\[0.5em]
f_{33}^*(t) &= f_{11}^*(t)+t+0.5.
\end{aligned}
\]	%
	These functions are presented in Figure~2 in the Supplementary Material.	It is worth noting that the relevant covariates for all clusters are	$X_{i1}, \ldots, X_{i6}$, each of which makes an equal contribution to	$Y_{i}$ since $\Vert f_{jk}\Vert _{2}=1$ for $j=1,\ldots,6$ and	$k=1,\ldots,K$. We set the number of clusters $K=3$.

	Next, we generate the cluster membership $Z_{i}$ from a multinomial distribution	as described in Section~\ref{sec:method}, with	$\pi _{k}=1/K, k=1,\ldots,K$. Providing	$\boldsymbol{X}_{i}(t),Z_{i}$, and $\boldsymbol{\beta }(t)$,	$Y_{i}(t)$ is generated from the model \eqref{eq: model} for each
	$t$, where the $\varepsilon _{ik}(t)$ is constructed as	$\tilde{\varepsilon}_{ik}(t)+\tau _{ik}$. For each $i$ and $k$,	$\{\tilde{\varepsilon}_{ik}(t);t\in [0,1]\}$ are independent mean-zero	Gaussian processes with the covariance function	${\sigma _{k}^{2}}\operatorname{exp}(-|t_{1}-t_{2}|)/2$, and $\tau _{ik}$s are Gaussian
	measurement noises with variance $\sigma _{k}^{2}/2$. The variance of the	error terms $\sigma _{k}^{2}$ is taken according to the signal-to-noise ratio (SNR)	%
	\begin{equation*}
		\sigma _{k}^{2}= \biggl\{ \frac{\sum_{i=1}^{n}\int _{0}^{1}\sum_{k=1}^{K}I(Z_{i}=k)  \{\boldsymbol{X}_{i}(t)^{\top}\boldsymbol{\beta }_{\cdot k}(t)  \}^{2}\,\mathrm{d}t}{n} \biggr\} /
		\text{SNR},
	\end{equation*}	%
	with SNR being set as 12. Finally, the observed time points	$\{t_{is};s=1,\ldots,S_{i}\}$ for functional data are set as 10 equally	spaced knots on $[0,1]$ for each $i$, mimicking the setting of our SDOH
	data.
	
	Based on the above setting, we evaluate the method proposed in Section~\ref{sec:method}, which is denoted as FGS-Net due to the use of FGS-Net	penalty \eqref{eq: penalty FGSCADL2}. We compare FGS-Net by other REM methods	penalized with the FS-Net penalty \eqref{eq: penalty FSCADL2} and FS penalty	\eqref{eq: hybrid norm}, denoted as FS-Net and FS in what follows. Apart	from these three methods, we examine the following competing methods for	clustering longitudinal associations:
	\begin{itemize}
		\item RP: This method incorporates a roughness penalty (RP) into the REM		algorithm, which is given as		$\sum_{j=1}^{p}\sum_{k=1}^{K}\lambda \Vert \beta _{jk}^{\prime \prime }\Vert _{2}^{2}$.
		\item VS-RP: This method initially employs the FGS-Net within the finite		mixture of functional linear concurrent model \eqref{eq: model}, setting		$K=1$ for variable selection (VS). This method is denoted as FMFLCM(1),		encompassing several existing methods		(\citet{wang2008variable,goldsmith2017variable,ghosal2020variable}) as special		cases. Subsequently, the preselected variables are used to implement the
		RP method for clustering.
		\item LI-MIX: This method fits the data by a finite mixture of linear regression		models (LI-MIX), that is, the functional coefficient		$\beta _{jk}(t)$ in \eqref{eq: model} is treated as a constant over		$t$. To preform this method, we first conduct a variable selection using		conventional linear regression models shrunk with an elastic-net penalty		(\citet{zou2009adaptive}). After that, we adopt the selected covariates to		fit a finite mixture of linear regression models for the clustering		(\citet{peel2000finite, khalili2007variable,khalili2013regularization}).
	\end{itemize}
	The RP is a simplification of the FS-Net penalty	\eqref{eq: penalty FSCADL2} with $\rho =0$, which only regularizes the	roughness of each $\beta _{jk}$ and does not yield sparsity. VS-RP and	LI-MIX are two two-step approaches, which implement the variable selection	and clustering orderly. These two-step methods are more simple than the	aforementioned FGS-Net, FS-Net, FS, and RP. However, selecting relevant	covariates prior to the clustering procedure may raise additional problems,	as the clustering performance may be sensitive to the outcome from the	variable selection. It is worth noting that the two-step method LI-MIX	further ignores the time-varying nature of $\beta _{jk}$.

	For each scenario with different combinations of $n$, $p$, and	$\alpha $, the simulations are repeated 100 times. We adopt random initialization	for the FGS-Net, FS-Net, and FS methods by setting	$\omega _{ik}^{(0)} = I(Z_{i}^{(0)} = k)$, where $Z_{i}^{(0)}$ is sampled	from a multinomial distribution over $\{1, \dots, K\}$ with equal probabilities.	On the other hand, RP, VS-RP, and LI-MIX are initialized with the actual	cluster membership $\omega _{ik}^{(0)}=I(Z_{i}=k)$. To alleviate computation	burdens, we only use one initialization for each simulation. Moreover,	the $K$ in RP, VS-RP, and LI-MIX is fixed to 3, the true number of clusters.	The $K$ for FGS-Net, FS-Net, and FS are selected based on	\eqref{eq:BIC}.
	
	The performance of clustering, variable selection, and parameter estimation	is evaluated based on the following criteria:	%
	\begin{itemize}		%
		\item Clustering accuracy is evaluated using the adjusted Rand Index (ARI, \citet{rand1971objective}). The ARI is bounded by $\pm 1$ to measures		the similarity between the true cluster membership and the estimated cluster		membership. A higher ARI represents a better clustering result.		%
		\item Variable selection performance is evaluated using C and IC, where		C is the number of zero coefficients that are correctly estimated to zero $\text{C}=\sum_{j=7}^{p}\sum_{k=1}^{K}I\bigl( \lVert \hat{\beta}_{jk}\rVert _{2}=0\bigr),$	%
		where $\hat{\beta}_{jk}$ is the estimate of $\beta _{jk}$. Similarly, IC		is the number of nonzero coefficients that are incorrectly estimated to		zero		%
		$\text{IC}=\sum_{j=1}^{6}\sum_{k=1}^{{K}}I\bigl( \lVert \hat{\beta}_{jk}\rVert _{2}=0\bigr).$	%
		\item The parameter estimation accuracy is measured using the standardized
		mean square error (MSE) of the functional coefficients: $\text{MSE}= \frac{\sum_{k=1}^{K}\sum_{j=1}^{p} \lVert  \beta _{jk}-\hat{\beta}_{jk} \rVert  _{2}^{2}}{\sum_{k=1}^{K}\sum_{j=1}^{p} \lVert  \beta _{jk} \rVert  _{2}^{2}}.$
	\end{itemize}

	\subsection{Result}

	We investigate the performances of different methods under various scenarios	of $n$, $p$, and $\alpha $. Here we set $n$ to $180$ or $300$ and take	$p$ as $10, n/6, n/2$, and $3n/4$, to consider the situations ranging	from a small to a large number of covariates. Additionally, we set	$\alpha $ to $0.4$ and $0.8$ to reflect the mild or strong dependence among	the covariates. The averaged ARI, C, IC, and MSE are presented in Table~\ref{tab:n180300}. In the analysis below, we only focus on the results	of $n=180$. Similar conclusions can be obtained for the result of
	$n=300$.
	
	Overall, FGS-Net, FS-Net, and FS demonstrate superior performance across	various scenarios with different values of $n$, $p$, and $\alpha $, underscoring	the benefits of simultaneously implementing variable selection and clustering	under the REM framework. In contrast, the RP method, which only applies	a roughness penalty and omits variable selection within the clustering	process, quickly deteriorates in performance for both clustering and parameter	estimation as $p$ increases. Among the two-step methods, we find that VS-RP	performs poorly compared to the first three REM-type methods. For instance,
	in all scenarios with $n=180$, the average ICs of VS-RP are mostly larger	than 9, indicating that about half of the nonzero functional coefficients	are incorrectly identified as zero. These results lead to significant estimation	errors in both clustering and parameter estimation by the VS-RP method,	suggesting that selecting variables prior clustering is ineffective. Notably,	the performance of LI-MIX is even poorer, as this method additionally ignore	the time-varying nature of $\boldsymbol{\beta }(\cdot )$ during the estimation	process.

\begin{table}[ht]
\centering
\scriptsize
\renewcommand{\arraystretch}{1}
\setlength\tabcolsep{2.4pt}
\caption{Averaged ARI, C, IC, and MSE for FGS-Net, FS-Net, FS, RP, VS-RP, and LI-MIX with $p=10, n/6,n/2,4n/3$, $\alpha=0.4,0.8$, $n=180,300$. The highest ARI, C, and lowest IC, MSE are in bold.}
{
\begin{tabular}{ll|llll|llll|llll|llll}
  \hline 
  \multicolumn{18}{c}{$n=180$}\\
  \hline
  &  &\multicolumn{4}{c|}{$p$=10} &\multicolumn{4}{c|}{$p$=30} &\multicolumn{4}{c|}{$p$=100} &\multicolumn{4}{c}{$p$=240} \\ 
  \hline
 
 & Model & ARI & C & IC & MSE & ARI & C & IC & MSE & ARI & C & IC & MSE & ARI & C & IC & MSE \\
 \hline

   \multirow{7}{*}{$\alpha=0.4$}   &  Truth & 1 & 12 & 0 & 0 & 1 & 72 & 0 & 0 & 1 & 252 & 0 & 0 & 1 & 702 & 0 & 0  \\ 
   & RP & \textbf{1.00} & 0.00 & \textbf{0.00} & 0.04 & 0.01 & 0.00 & \textbf{0.00} & 1.00 & 0.01 & 0.00 & \textbf{0.00} & 1.00 & 0.01 & 0.00 & \textbf{0.00} & 1.00 \\ 
   & VS-RP & 0.49 & 11.94 & 7.59 & 0.60 & 0.46 & 71.61 & 7.56 & 0.61 & 0.37 & 250.86 & 7.50 & 0.65 & 0.30 & 699.00 & 6.90 & 0.70 \\ 
   & LI-MIX & 0.28 & 4.89 & 2.58 & 0.72 & 0.22 & 36.30 & 2.82 & 0.81 & 0.07 & 139.11 & 2.85 & 1.24 & 0.01 & 388.71 & 3.09 & 2.76 \\ 
   & FS & \textbf{1.00} & 11.99 & \textbf{0.00} & \textbf{0.02} & 0.99 & 71.99 & 0.12 & \textbf{0.02} & 0.94 & 251.80 & 0.97 & 0.13 & 0.87 & 701.73 & 2.18 & 0.24 \\  
   & FS-Net & \textbf{1.00} & \textbf{12.00} & \textbf{0.00} & \textbf{0.02} & 0.99 & \textbf{72.00} & 0.12 & \textbf{0.02} & 0.94 & 251.84 & 1.00 & 0.12 & 0.87 & 701.90 & 2.28 & 0.24 \\ 
   & FGS-Net & \textbf{1.00} & \textbf{12.00} & \textbf{0.00} & \textbf{0.02} & \textbf{1.00} & \textbf{72.00} & \textbf{0.00} & \textbf{0.02} & \textbf{0.99} & \textbf{252.00} & \textbf{0.00} & \textbf{0.06} & \textbf{0.97} & \textbf{702.00} & 0.33 & \textbf{0.08} \\

      \hline
     \multirow{7}{*}{$\alpha=0.8$}& Truth & 1 & 12 & 0 & 0 & 1 & 72 & 0 & 0 & 1 & 252 & 0 & 0 & 1 & 702 &0  & 0 \\ 
     & RP & \textbf{1.00} & 0.00 & \textbf{0.00} & 0.09 & 0.01 & 0.00 & \textbf{0.00} & 0.99 & 0.01 & 0.00 & \textbf{0.00} & 1.00 & 0.01 & 0.00 & \textbf{0.00} & 1.00 \\ 
     & VS-RP & 0.77 & 11.49 & 10.41 & 0.88 & 0.74 & 70.92 & 10.50 & 0.89 & 0.73 & 250.44 & 10.20 & 0.87 & 0.66 & 699.24 & 10.62 & 0.90 \\ 
     & LI-MIX & 0.23 & 6.06 & 3.96 & 0.88 & 0.19 & 43.44 & 4.32 & 1.00 & 0.08 & 145.47 & 4.47 & 1.98 & 0.01 & 426.78 & 5.10 & 4.93 \\ 
     & FS & 0.99 & 11.91 & 1.86 & 0.21 & 0.99 & 70.62 & 2.23 & 0.25 & 0.99 & 249.03 & 3.39 & 0.34 & 0.96 & 701.88 & 8.51 & 0.77 \\ 
     & FS-Net & 0.99 & 11.90 & 0.53 & 0.11 & 0.99 & 71.04 & 0.61 & 0.12 & 0.99 & 249.55 & 1.23 & 0.16 & 0.97 & 701.87 & 6.71 & 0.55 \\ 
     & FGS-Net & \textbf{1.00} & \textbf{12.00} & 0.15 & \textbf{0.09} & \textbf{0.99} & \textbf{72.00} & 0.27 & \textbf{0.11} & \textbf{0.99} & \textbf{252.00} & 0.36 & \textbf{0.13} & \textbf{0.98} & \textbf{701.97} & 1.08 & \textbf{0.17} \\

\hline 
  \multicolumn{18}{c}{$n=300$}\\
  \hline
     &  &\multicolumn{4}{c|}{$p$=10} &\multicolumn{4}{c|}{$p$=50} &\multicolumn{4}{c|}{$p$=150} &\multicolumn{4}{c}{$p$=400} \\ 
  \hline
  & Model & ARI & C & IC & MSE & ARI & C & IC & MSE & ARI & C & IC & MSE & ARI & C & IC & MSE \\ 
  \multirow{7}{*}{$\alpha=0.4$}  & Truth & 1 & 12 & 0 & 0 & 1 & 132 & 0 & 0 & 1 & 432 & 0 & 0 & 1 & 1182 & 0 & 0  \\ 
  & RP & \textbf{1.00} & 0.00 & \textbf{0.00} & 0.04 & 0.01 & 0.00 & \textbf{0.00} & 1.00 & 0.01 & 0.00 & \textbf{0.00} & 1.00 & 0.00 & 0.00 & \textbf{0.00} & 1.00 \\
  & VS-RP & 0.88 & 11.82 & 3.33 & 0.25 & 0.87 & 131.22 & 3.42 & 0.25 & 0.88 & 430.11 & 3.54 & 0.26 & 0.85 & 1177.29 & 3.42 & 0.28 \\ 
   
  & LI-MIX & 0.31 & 4.77 & 1.26 & 1.14 & 0.25 & 65.34 & 1.65 & 1.19 & 0.09 & 232.98 & 1.59 & 1.42 & 0.01 & 632.7 & 1.59 & 2.99 \\ 
  
  & FS & \textbf{1.00} & \textbf{12.00} & \textbf{0.00} & \textbf{0.01} & \textbf{0.99} & 131.99 & 0.08 & 0.03 & 0.99 & 431.98 & 0.06 & 0.03 & \textbf{0.99} & 1181.93 & 0.15 & 0.03 \\ 
  & FS-Net & \textbf{1.00} & \textbf{12.00} & \textbf{0.00} & \textbf{0.01} & 0.99 & \textbf{132.00} & 0.10 & 0.03 & 0.99 & \textbf{432.00} & 0.07 & 0.03 & \textbf{0.99} & 1181.97 & 0.16 & 0.03 \\ 
  & FGS-Net & \textbf{1.00} & \textbf{12.00} & \textbf{0.00} & \textbf{0.01} & \textbf{1.00} & \textbf{132.00} & \textbf{0.00} & \textbf{0.01} & \textbf{1.00} & \textbf{432.00} & \textbf{0.00} & \textbf{0.01} & \textbf{0.99} & \textbf{1182.00} & \textbf{0.00} & \textbf{0.02} \\ 
     [1pt]
     
  \hline
  \multirow{7}{*}{$\alpha=0.8$}  & Truth & 1 & 12 & 0 & 0 & 1 & 132 & 0 & 0 & 1 & 432 & 0 & 0 & 1 & 1182 & 0 & 0  \\
  & RP & \textbf{1.00} & 0.00 & \textbf{0.00} & 0.07 & 0.01 & 0.00 & \textbf{0.00} & 0.99 & 0.01 & 0.00 & \textbf{0.00} & 1.00 & 0.01 & 0.00 & \textbf{0.00} & 1.00 \\   
  & VS-RP & 0.92 & 11.73 & 8.04 & 0.7 & 0.91 & 130.92 & 7.89 & 0.7 & 0.9 & 429.57 & 7.74 & 0.68 & 0.85 & 1175.85 & 8.13 & 0.69 \\  
  & LI-MIX & 0.24 & 6.42 & 2.88 & 1.16 & 0.2 & 80.13 & 2.91 & 1.26 & 0.08 & 263.16 & 3.3 & 1.82 & 0.02 & 747.66 & 3.57 & 4.57 \\ 
  & FS & \textbf{1.00} & \textbf{12.00} & \textbf{0.00} & \textbf{0.04} & \textbf{1.00} & 131.65 & \textbf{0.00} & 0.08 & 0.99 & 430.79 & 0.09 & \textbf{0.04} & 0.97 & 1181.97 & 8.02 & 0.72 \\ 
  & FS-Net & \textbf{1.00} & 11.99 & \textbf{0.00} & \textbf{0.04} & \textbf{1.00} & 131.74 & \textbf{0.00} & \textbf{0.04} & 0.99 & 431.05 & 0.07 & \textbf{0.04} & 0.98 & 1181.96 & 6.14 & 0.49 \\ 
  & FGS-Net & 0.99 & \textbf{12.00} & \textbf{0.00} & \textbf{0.04} & 0.99 & \textbf{132.00} & 0.06 & \textbf{0.04} & \textbf{1.00} & \textbf{432.00} & \textbf{0.00} & \textbf{0.04} & \textbf{0.99} & \textbf{1182.00} & \textbf{0.30} & \textbf{0.09} \\ 
   \hline

\end{tabular}
}
\label{tab:n180300}
\end{table}

	In Table~\ref{tab:n180300} we observe that the ICs and MSEs of FS are	significantly larger than those of FS-Net as $\alpha $ increases. This	is expected since the dependencies between functional covariates may impede	the performance of the FS procedure. Incorporating an additional ridge-type	penalty in FS-Net could help stabilize the estimation process for both	variable selection and parameter estimation. Moreover, as $p$ increases,	the ICs and MSEs of FS-Net become further larger than those of FGS-Net. 	In these scenarios, the high dimensionality may undermine the statistical	efficiency for both variable selection and clustering in FS-Net. Therefore,	it would be beneficial to impose cluster-invariant sparsity through FGS-net
	to borrow strength across all clusters during estimation.

	\begin{figure}[ht]
		\centering
		\includegraphics[width = 1\linewidth]{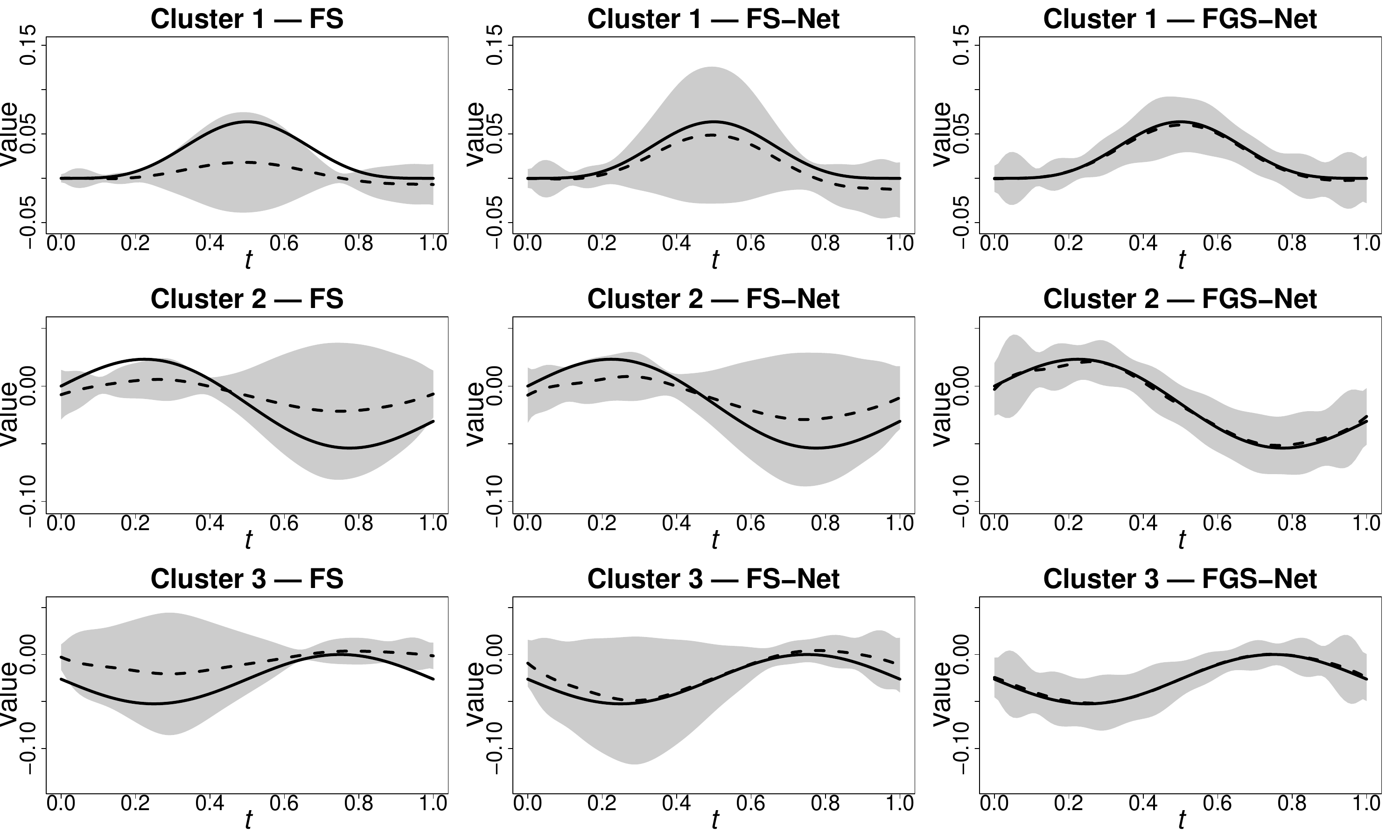}
		\vspace{-0.6cm}
		\caption{
			The true curves of the functional coefficients $\beta_{jk}$ from each cluster $k$ and their estimations from FS, FS-Net, and FGS-Net, for $n=180$, $p=240$, and $\alpha=0.8$. The solid black lines represents the true curve, while the dashed lines, along with their corresponding bands, represent the pointwise means and the 95\% confidence bands over 100 replicated simulations.
		}
		\label{fig: simucoef}
	\end{figure}

	In the case of $n=180$, $p=240$, and $\alpha =0.8$, we further illustrate	the estimation performance of the functional coefficients in Figure~\ref{fig: simucoef}. We observe that the estimations of FS-Net exhibit
	smaller biases, compared to FS, highlighting the significance of FS-Net	in mitigating biases of functional coefficients. In comparison to the results	of FS-Net and FS, the estimated curves of FGS-Net show even smaller biases	and narrower confidence bands, owing to the pursuit of cluster-invariant	sparsity.

	In Section C.4 of the Supplementary Material, we assess the computational 	cost of FGS-Net in comparison to the two-step methods VS-RP and LI-MIX. 	Our findings indicate that, while the two-step methods are computationally	faster than those utilizing REM algorithms, their performance is generally
	inferior, as shown in Table~\ref{fig: simucoef}, due to the uncorrelatedness	of the variable selection and clustering steps. Additionally, we observe	that the computational cost of FGS-Net does not increase too rapidly as	$n$ and $p$ increase. This suggests tolerable computational costs for simultaneously	clustering and variable selection in longitudinal associations for large	datasets.

	\section{Real data}
	\label{sec: realdata}

	In this section we apply our methods to the SDOH and stroke mortality dataset	for clustering their longitudinal associations. Given that the stroke mortality	data are right-skewed and take positive values, we conduct a log transformation	to stabilize its variance. Besides, we normalize the data	$X_{ij}(t_{S_{i}})$ by
	$({\sum_{i} S_{i}})^{-1}\sum_{i=1}^{n}\sum_{s=1}^{S_{i}}X_{ij}(t_{S_{i}})^{2}$	to ensure that the SDOH covariates are on the same scale.

	\paragraph*{Performance Comparison}
	
	We apply the proposed FS-Net and FGS-Net to the dataset and compare these	methods with FMFLCM(1)	(\citet{wang2008variable,goldsmith2017variable,ghosal2020variable}) and the	LI-MIX method	(\citet{peel2000finite,khalili2007variable,khalili2013regularization}) in	the simulation section. Notably, FMFLCM(1) is a special case of the FMFLCM	in \eqref{eq: model} with $K=1$, indicating that it does not cluster longitudinal	associations during variable selection. Furthermore, LI-MIX does not account
	for longitudinal changes in the SDOH-stroke mortality associations during	clustering, and FS-Net focuses only on cluster-variant sparsity in model	estimation. Detailed comparisons of these methods are presented in Section	C.3 of the Supplementary Material. Overall, FMFLCM(1) and LI-MIX tend to	select a large number of covariates but fail to adequately fit the data.	In contrast, FS-Net and FGS-Net provide a more accurate fit, highlighting	the importance of accommodating longitudinal changes and heterogeneity
	when analyzing associations between SDOH and stroke mortality.

	It is worth noting that the model in FS-Net has a higher complexity, compared	to that of FGS-Net, with the latter pursuing cluster-invariant sparsity	rather than cluster-variant sparsity. In our analysis we find that both
	methods achieved the fitting residuals of 0.15, and their performances	are quite similar in terms of clustering and variable selection. These	findings suggest that the increased complexity of FS-Net does not result	in significant improvements in model fitting. For a more interpretable	result, we focus only on the data analysis of FGS-Net in the remaining	and refer to Section C.3 of the Supplementary Material for the results	of FS-Net.

	\paragraph*{Data Analysis}
	
\begin{figure}[ht]
		\centering
		\includegraphics[width=0.9\linewidth]{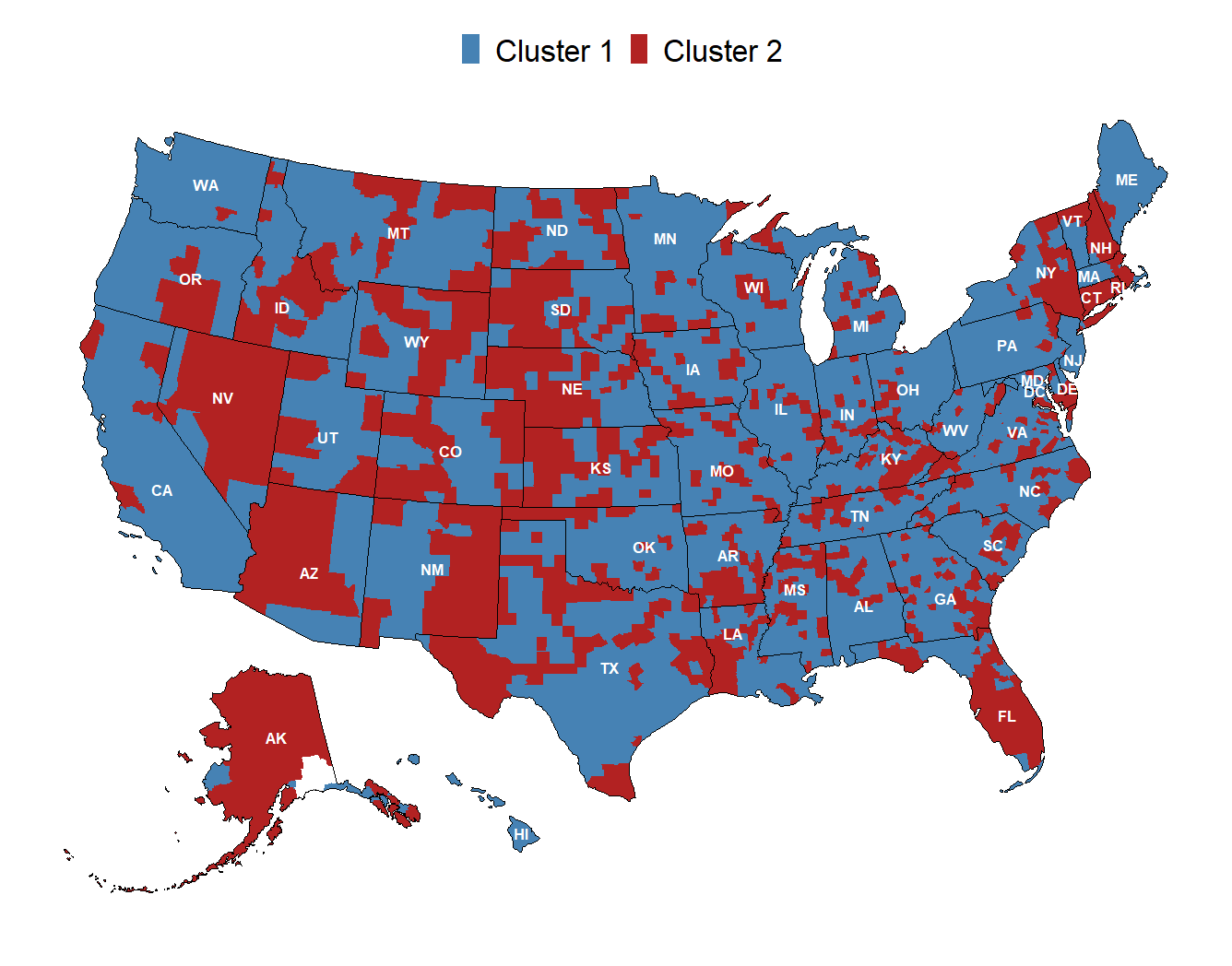}
		\vspace{-0.5cm}
		\caption{
			The clustering result for the county-level longitudinal associations between the SDOH and stroke mortality in the US.}
		\label{fig: clusterres}
	\end{figure}

	Using FGS-Net, we identify two clusters for the longitudinal associations	and 18 relevant SDOH covariates for stroke mortality. The two clusters	for the county-level longitudinal associations are presented in Figure~\ref{fig: clusterres}. The proportions of two clusters, determined by the	number of counties, stand at 68\% and 32\%, respectively. Notably, both	clusters are prevalent across the majority of states in the U.S., encompassing	both rural and urban areas (urban: 76\% and 24\%, and rural: 65\% and 35\%,	for cluster 1 and cluster 2, respectively). It is worth noting that the	southeastern U.S. contains a region called the Stroke Belt	(\citet{heyden1978coffee,lanska1995geography,karp2016reassessing,howard2020twenty}),	known for its persistent high relative excess of stroke mortality. Despite	counties in the Stroke Belt having similar stroke severity, this area is	also mixed by the two clusters, with proportions of 70\% and 30\%, respectively.	These results suggest that regions sharing similar geographic and stroke	characteristics may have very different SDOH-stroke mortality associations,	and we may need to consider separating two types of policies for the SDOH	adjustments in stroke management based on our clustering results.

	In addition, we illustrate the selected 18 covariates of SDOH in Table~\ref{tab: sdohvar}, ordered by their relative importance	(\citet{gromping2007relative}), defined as	%
	\begin{equation*}
		\text{RI}(\boldsymbol{\beta }_{j\cdot})= \lVert \boldsymbol{\beta}_{j\cdot} \rVert _{2} \Biggl\{\frac{1}{n}\sum_{i=1}^{n} \Biggl\lVert X_{ij}-		\frac{1}{n}\sum_{i=1}^{n}X_{ij}
		\Biggr\rVert ^{2} \Biggr\}^{1/2}.
	\end{equation*}	%
	We find that the influence of the SDOH on stroke mortality is mainly contributed	by four aspects: social environment, built environment, health care system,	and biology. Beyond the well-studied determinants from economic, cultural,	and racial domains (\citet{tsao2023heart}), we find that stroke mortality	is significantly associated with living and working environments, education	level, and overuse of opioids. Typically, among the selected variables	of SDOH, most of them are related to economic development. For example,	in Table~\ref{tab: sdohvar}, MEDIAN\_HOME\_VALUE may reflect overall economic	development and infrastructure building in the community. Additionally,	a higher value of ELDERLY\_RENTER may suggest a larger elderly population	with lower income, facing issues such as housing instability. These economic-related	factors may be potentially adjusted through more equitable economic policies.	For example, MEDIAN\_HOME\_VALUE can be adjusted by facilitating economic
	development. In addition, serving as a sign of the elderly living condition,	ELDERLY\_RENTER can be adjusted by improving elderly welfare.
	
	\begin{table}[ht]
		\centering
		\caption{{The selected SDOH variables ranked by descent order of their relative importance.}}
		\scriptsize
		\begin{tabu}{|l|X[l]|l|}
			\hline
			Variable & Explanation & SDOH Domain \\ 
			\hline
			MEDIAN\_HOME\_VALUE & Median home value of owner-occupied housing units & Sociocultural Environment \\  \hline 
			NO\_ENGLISH & Population that does not speak English at all, \% & Sociocultural Environment \\   \hline
			ASIAN\_LANG & Population that speaks Asian and Pacific Island languages, \% & Sociocultural Environment \\   \hline
			ELDERLY\_RENTER & Rental units occupied by householders aged 65 and older, \% & Built Environment \\   \hline
			GINI\_INDEX & Gini index of income inequality & Sociocultural Environment \\   \hline
			HOME\_WITH\_KIDS & Owner-occupied housing units with children, \% & Built Environment \\   \hline
			AGRI\_JOB & Employed working in agriculture, forestry, fishing and hunting, and mining, \% & Sociocultural Environment \\   \hline
			NO\_VEHICLE & Housing units with no vehicle available, \% & Built Environment \\   \hline
			OPIOID & Number of opioid prescriptions per 100 persons & Health Care System \\   \hline
			WEAK\_ENGLISH & Population that does not speak English well, \% & Sociocultural Environment \\   \hline
			BLACK & Population reporting Black race, \% & Biology \\   \hline
			BACHELOR & Population with a bachelor's degree, \% & Sociocultural Environment \\   \hline
			NO\_FUEL\_HOME & Occupied housing units without fuel, \% & Built Environment \\   \hline
			TIME & Time Effect & Sociocultural Environment \\   \hline
			ONLY\_ENGLISH & Population that only speaks English, \% & Sociocultural Environment \\   \hline
			MOBILE\_HOME & Housing units that are mobile homes, \% & Built Environment \\   \hline
			BELOW\_HIGH\_SCH & Population with less than high school education, \% & Sociocultural Environment \\   \hline
			DRIVE\_2WORK & Workers taking a car, truck, or van to work, \% & Built Environment \\ 
			\hline
		\end{tabu}
		\label{tab: sdohvar}
	\end{table}

	\begin{figure}[ht]
		\centering
		\includegraphics[scale = 0.13]{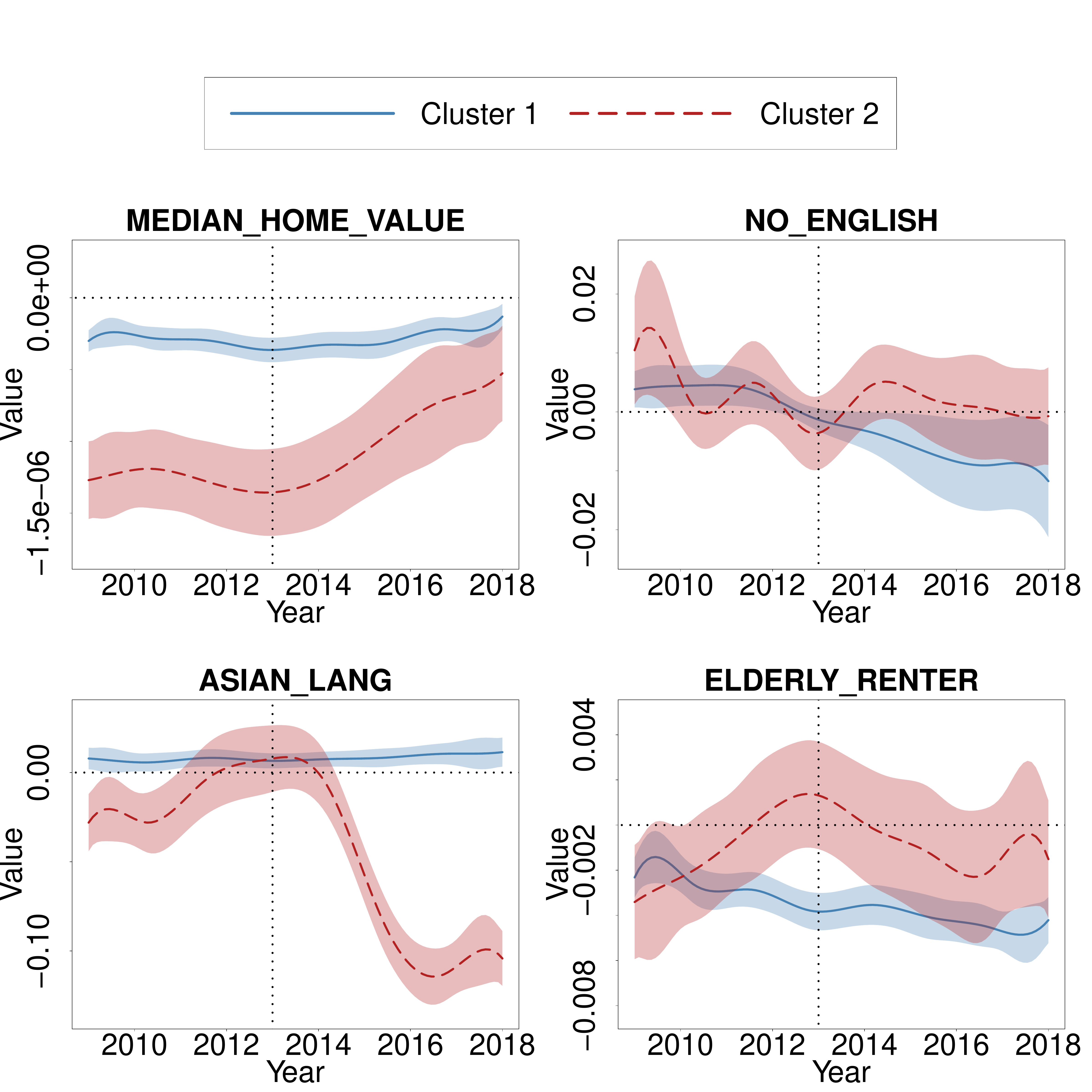}
		\vspace{-0cm}
		\caption{The estimated longitudinal associations for the leading four SDOH in two clusters. The dotted horizontal line in each sub-figure represents the zero line for the SDOH-stroke mortality longitudinal associations, and the dotted vertical line in each sub-figure indicates the common reflection point of the longitudinal associations in cluster 2.
		}
		\label{fig: coefrd}
	\end{figure}
	
	To further quantify the longitudinal associations between SDOH and stroke	mortality, we investigate the functional coefficients $\beta _{jk}$ estimated	from the dataset. Here we only focus on the coefficients from the leading	four important SDOH, which are MEDIAN\_HOME\_VALUE, NO\_ENGLISH, ASIAN\_LANG,	and ELDERLY\_RENTER. The longitudinal associations in the two clusters,
	along with the 95\% confidence bands, are shown in Figure~\ref{fig: coefrd}. These 95\% confidence bands are constructed using a	wild-bootstrap procedure similar to those in	\cite{zhu2012multivariate}. The detailed bootstrap procedure and its effectiveness	are demonstrated in Section C.2 of the Supplementary Material.

	In Figure~\ref{fig: coefrd} we observe that the leading four SDOH possess	time-varying associations with stroke mortality. Furthermore, these associations	are notably different between the two clusters. For example, the fluctuation	of the longitudinal associations in cluster 2 is more significant than	those in cluster 1. Additionally, we observe that, in cluster 2, all four	associations exhibit a common inflection point in 2013 (the dotted vertical	lines in Figure~\ref{fig: coefrd}). These findings help us better understand
	the dynamics of the associations between SDOH and stroke mortality in the U.S..

	In Figure~\ref{fig: coefrd}, MEDIAN\_HOME\_VALUE shows a stable and weak	association with stroke mortality for the counties in cluster 1, while	those in cluster 2 exhibit a more intense negative effect with stroke mortality.	The situation for ELDERLY\_RENTER differs with the association in cluster	2 showing significant fluctuation, while that in cluster 1 presents a time-increasing	negative association with stroke mortality.

	As economic-related factors, MEDIAN\_HOME\_VALUE and ELDERLY\_RENTER have	been identified to significantly associate with stroke mortality	(\citet{rodgers2019county, mawhorter2023housing}). Our findings not only	align with existing literature but also provide further insights into more	tailored stroke mortality prevention strategies related to SDOH variables.	For instance, the notable negative association between MEDIAN\_HOME\_VALUE	and stroke mortality in cluster 2 highlights the importance of focusing
	on infrastructure development, maintenance, and enhancing overall economic	equity in these regions. Noting the time-increasing negative association	between ELDERLY\_RENTER and stroke mortality in cluster 1, it may be beneficial	to prioritize viable housing programs for lower-income elderly populations
	in these regions. 
	
	In addition, NO\_ENGLISH and ASIAN\_LANG are SDOH variables that measure	the sociocultural environment of the county. NO\_ENGLISH, which indicates	the percentage of the population that does not speak English, is associated	with stroke mortality in both clusters, with the direction of the associations
	shifting from positive to negative. A similar decreasing trend is observed	in the association between ASIAN\_LANG (measuring the percentage of the	population that speaks Asian and Pacific Island languages) and stroke mortality	in cluster 2. Conversely, the association between ASIAN\_LANG and stroke
	mortality in cluster 1 remains stable, exhibiting a weakly positive trend	over time.

	The above results suggest that the association between the density of immigrants	or Asian and Pacific Islanders and stroke mortality may not be uniform	across regions and time periods. It is worth considering that immigrants	or Asian and Pacific Islanders who experience a stroke may encounter challenges	in accessing timely stroke care if they do not speak English. This language	barrier may potentially contribute to an increased risk of stroke-related	death in specific regions and during certain time periods	(\citet{shah2015impact}). However, our findings indicate that their risks	with stroke-related death may have diminished from 2010 to 2018. This reduction	may be caused by gradual improvements in stroke care for immigrants or	Asian and Pacific Islanders in the U.S., particularly in the counties of	cluster 2.

	\section{Dicussion}
	\label{sec: discussion}

	In this article, we introduce a novel clustering method for regional divisions	of U.S. counties based on their longitudinal associations between SDOH	and stroke mortality. The challenges for this task arise from the latent	and cluster-specific nature of the associations, which are compounded by	their functional and high-dimensional characteristics simultaneously. To	tackle these complex structures, we propose an REM algorithm that utilizes	a finite mixture of functional linear concurrent models. Our method explores
	the clustering-invariant sparsity via a	FGS-Net penalty within the REM algorithm, allowing for efficient variable	selection in functional covariates to identify the most significant associations	among all clusters. The effectiveness of our method has been demonstrated	via extensive simulations. In the end we apply the proposed method to the	SDOH--stroke mortality longitudinal data, facilitating the regional divisions
	of U.S. counties. This enables the identification of regions for informing	SDOH--targeted prevention of stroke mortality.
	
	In the analysis of the SDOH and stroke mortality dataset, our clustering	map in Figure~\ref{fig: clusterres} provides a novel result for region-division	in stroke management, taking into account the similarity among longitudinal	associations between SDOH and stroke mortality. These findings indicate	that heterogeneity in these associations occurs even within areas that	share similar geographical conditions, stroke mortality rates, or economic	statuses. Therefore, region divisions based solely on these factors may	not be effective for SDOH-based stroke death control. Moreover, we uncover	various patterns of longitudinal associations among the two identified	clusters of U.S. counties. The dynamics within these associations, including	scale, trends, and inflection points, not only provide heuristic information	for understanding complex SDOH-stroke mortality associations but also are	useful for establishing timely SDOH adjustments in stroke death control.
	
	Our method can be generalized to other clustering tasks, such as time-varying	clustering (\citet{sartorio2020dynamic}) and spatial smooth clustering	(\citet{li2019spatial,liang2021modeling}). In these applications, we can relax	the assumption of fixed cluster membership to accommodate the desired clustering	structures. For instance, to implement time-varying clustering, we can	allow cluster memberships to change over time within the finite mixture	of functional linear concurrent models; for spatial smooth clustering,	spatial dependence structures can be integrated into the model of cluster	memberships (\citet{jiang2012clustering,liang2021modeling}). All these extensions	can be achieved by modifying the likelihood function in the REM algorithm,	tailored to specific research problem of interest. We leave these as future	directions for investigation.

	\small
	\bibliography{ref}
	\bibliographystyle{chicago}    
\end{document}